\documentclass{iopart}

\usepackage{graphicx,amsfonts,amsbsy,amssymb,amsthm}
\usepackage{color}
\usepackage{epstopdf}
\usepackage{hyperref}
\usepackage{cite}

\newcommand{\E}{\mathbf{E}}
\newcommand{\Erf}{{\rm Erf}}
\newcommand{\crit}{{\rm crit}}

\newcommand{\Var}{{\rm Var}}

\renewcommand{\choose}[2]{\left( \begin{array}{c} #1 \\ #2 \end{array}\right)}
\newcommand{\bsigma}{\boldsymbol{\sigma}}
\renewcommand{\bs}{\boldsymbol{s}}

\renewcommand{\O}[1]{\mathcal{O}(N^{#1})}

\begin{document}

\title[A random walk approach for Bernoulli ensembles]{Spectral statistics of Bernoulli matrix ensembles - a random walk approach (I)}

\author{Christopher H. Joyner$^{1,2}$  and Uzy Smilansky$^{1}$}
\address{$^1$Department of Physics of Complex Systems,
Weizmann Institute of Science, Rehovot 7610001, Israel.}
\address{$^{2}$School of Mathematical Sciences, Queen Mary University of London, London, E1 4NS, UK}
 \ead{\mailto{christopher.joyner@weizmann.ac.il}, \mailto{uzy.smilansky@weizmann.ac.il}}
%PACS 05.45.Mt , 02.10.Ox %
\begin{abstract}
We investigate the eigenvalue statistics of random Bernoulli matrices, where the matrix elements are chosen independently from a binary set with equal probability. This is achieved by initiating a discrete random walk process over the space of matrices and analysing the induced random motion of the eigenvalues - an approach which is similar to Dyson's Brownian motion model but with important modifications. In particular, we show our process is described by a Fokker-Planck equation, up to an error margin which vanishes in the limit of large matrix dimension. The stationary solution of which corresponds to the joint probability density function of certain well-known fixed trace Gaussian ensembles.
\end{abstract}

\section{Introduction}

Random matrices were first introduced by Wishart in the 1930s in the context of multivariate statistical analysis \cite{Wishart}. However it was not until the late 1950s and early 1960s that random matrix theory (RMT) was significantly developed by Wigner, Dyson, Mehta and others \cite{Wigner,Dyson1,Mehta-1960}. This interest was initiated by Wigner's observation that eigenvalues of large random matrices and scattering resonances of large atomic nuclei shared the same statistical properties \cite{Wigner}. RMT has progressed in a similar vain ever since, as the development of new methods has paralleled the application to many new areas of mathematics and physics, including, for example, quantum chaos, disordered systems, number theory, field theory and financial mathematics \cite{Oxford-RMT}. One of the main reasons for this outstanding success has been the observation of universal features displayed across a wide range of systems, allowing RMT to predict the properties of many emergent spectral phenomena, regardless of the intrinsic microscopic details of the system. Understanding the reasons for this universality has been, and still is, one of the prime goals in RMT and related fields. Much progress has been made both in the theoretical physics community (see e.g. \cite{Oxford-RMT,Haake-2010,Guhr-1998} and references therein) and in mathematics (see e.g. \cite{Oxford-RMT,forrester} and references therein).

Recently, significant progress has also been made in proving universality for Wigner matrices - commonly referred to as the Wigner-Dyson-Mehta conjecture (p.7 of \cite{Mehtabook}, see also \cite{Tao-2011a,Erdos-2012}). In essence, this states that for large matrices, in which the elements are independent random variables, the distribution of eigenvalues should converge to the well known Gaussian ensembles, regardless of the precise distribution of the matrix elements. Two complementary approaches have been developed in this respect. The first, based upon the initial ideas of Johansson \cite{Johansson-2001}, is the so-called \emph{heat-flow} method, which analyses the stochastic motion of eigenvalues under a small Gaussian convolution \cite{Erdos-2010b,Erdos-2010c,Erdos-2012c,Erdos-2012d}. The other compares the eigenvalue statistics under the element-wise swapping of distributions in a random matrix \cite{Tao-2011b,Tao-2010,Tao-2012}.

%Extremely powerful tools, based around local semicircle estimates \cite{Erdos-2009a,Erdos-2010,Erdos-2012a,Erdos-2012b}, have facilitated a series of works culminating in essentially

Remarkably, this universality extends to Wigner matrices in which the entries are randomly chosen from a set consisting of just two elements \cite{Tao-2011b,Erdos-2011}, so-called \emph{Bernoulli ensembles}. The methods have also been adapted to treat sparse random matrices and distributions with non-zero mean \cite{Erdos-2013,Erdos-2012e}. However all these results rely on the independence of the matrix elements and a method for relaxing this condition for Bernoulli matrices is thus far out of reach. Although universality results have been shown in other classes of correlated matrices using orthogonal polynomial techniques \cite{Pastur-1997,Deift-1999,Bleher-1999}.

\vspace{10pt}

\noindent One of the principle reasons for studying the spectral properties of Bernoulli matrices is their inherent relation to graphs, or networks, since this binary choice can be used to represent the existence/absence of a connection between vertices, or nodes. These random graphs are used extensively in areas such as computer science, network modelling, statistical physics and neural networks to name but a few and analysis is often based upon their spectral properties. Therefore, a comprehensive understanding of this universality is of central importance. Often this binary choice is given by the set $\{0,1\}$, in which case the matrix is referred to as the \emph{adjacency matrix} of the graph.

By imposing different constraints on the matrices one can create ensembles of random graphs, that, by extension, can serve to characterise these Bernoulli ensembles. Some examples of adjacency matrices associated to well-studied types of graphs are listed below.

\begin{enumerate}

\item \label{Undirected graphs} {\it Undirected graphs} are given by symmetric adjacency matrices $A$ in which the elements $A_{ij} = A_{ji}$ are equal to 1 if vertices $i$ and $j$ are connected and 0 otherwise.

\item \label{Directed graphs} {\it Directed graphs} occur when the connection from vertices $i$ to $j$ is independent of the connection from $j$ to $i$. In this case $A_{ij}$ is 1 if there is a directed edge from $i$ to $j$ and 0 otherwise.

\item \label{Tournament graphs} {\it Tournament graphs} are unidirectional graphs, meaning all pairs of vertices $(i,j)$ are connected with an edge admitting a certain orientation. The matrix elements thus satisfy $A_{ji}=1-A_{i,j} \in \{0,1\}$.

\item {\it $d$-regular graphs} are undirected graphs in which every vertex has exactly the same degree, or valency, equal to $d$. Therefore each $A$ satisfies $\sum_j A_{ij} =d, \ \forall \ i$.

\item {\it Regular tournament graphs} are tournament graphs with an odd number of vertices $N$, in which each vertex $i$ has an equal number, $\frac{N-1}{2}$, of directed edges pointing towards and away from $i$. In other words $\sum_j A_{ij}= \frac{N-1}{2}, \ \forall \ i$.
\end{enumerate}

An ensemble of random graphs is then created by endowing a probability measure on the set of all graphs. The most common choice is the uniform distribution, so that each graph is equally likely to be selected. In the first three examples above, this equates to having \emph{iid} random variables for the matrix elements in which 0 and 1 are chosen with probability $1/2$, without any further constraint.  We shall therefore refer to these ensembles as \emph{unconstrained} and those with non-independent matrix elements, such as random $d$-regular graphs and random regular tournaments, as \emph{constrained}. In the present manuscript we shall be concerned solely with the former.

Returning to the adjacency matrices, then all the ensembles outlined above have been shown, \cite{Erdos-2013,Erdos-2012e}, or are conjectured \cite{Oren-2010,Miller-2008}, to display local eigenvalue statistics consistent with RMT, after appropriate rescaling. However these matrices possess two properties which render this comparison less than straightforward. The first is that $\mathbb{E}[A_{ij}]\ne 0$, in contrast with the Gaussian case in which the mean is zero. The second is that by the Perron-Frobenious theorem, the largest eigenvalue of the adjacency matrices is often clearly separated from the rest of the spectrum. We therefore define for all $A$
\begin{equation}\label{eq:balanced}
B= 2 A - J +I
\end{equation}
where $J$ denotes the matrix in which every element is 1 and $I$ is the identity. $B$ is the ``balanced" counterpart of $A$ and it does not suffer from the deficiencies mentioned above. But for a trivial shift and a factor $2$, the difference between $A$ and $B$ is a matrix of rank $1$, which ensures that the spectral properties of the $A$ and $B$ ensembles are closely related.

\vspace{10pt}

\noindent In a series of papers, of which this is the first, we introduce a random walk approach to the spectral statistics of Bernoulli ensembles. The main purpose of the present paper is to introduce the approach and apply it to various unconstrained ensembles. We are inspired by Dyson's Brownian motion model \cite{Dyson-1962}, which returns the joint probability density function (JPDF) for the eigenvalues of the canonical Gaussian ensembles as the stationary solution of an induced Fokker-Planck equation. However Dyson's model has a particular drawback - it is inextricably linked to the Gaussianity of the matrix elements. Here we overcome this restriction, obtaining the same Fokker-Planck equation as Dyson, but at the expense of an error caused by having a discrete time step dependent on the matrix size $N$. Therefore, whilst Dyson's model works exactly for any matrix size $N \geq 2$, our approach only becomes valid when $N\gg 1$. This relies on taking a series of discrete time steps and showing the linearity in time of the first two moments of the evolution kernel - as one would expect from arguments pertaining to the central limit theorem.

The paper is organised as follows: In Section \ref{random walks} we set up the discrete random walk process in the space of matrices and show it is equivalent to a random walk on the hypercube. The latter has been well studied previously \cite{Kac,Letac,Diaconis} and is known to be ergodic with a uniform stationary distribution. We are interested in how this process induces a random walk in a restricted space of variables (in our case the spectrum) and the transition from a discrete to a limiting continuous space-time description. Therefore, for the purposes of clarity, we first provide a simple example of this limiting transition in Section \ref{entropic forces}. This is the time dependent probability distribution of the walker to be at a given distance from the origin \cite{Kac,Letac,Diaconis}, which converges to the familiar (continuous) Ornstein-Uhlenbeck process in the limit of large hypercubes. This shows how entropic forces may arise from a simple reduction of variables and the same ideology will be used in order to investigate the spectral statistics of Bernoulli matrix ensembles in the remainder of the paper.

Our main result is outlined in Section \ref{Spectral statistics}. Here, using the method of Kolmogorov \cite{Kolmogorov} (see also \cite{Wang}), we study the evolution of observables under the random walk in the space of spectra. This leads to a Fokker-Planck equation with an error that decreases like $\O{-\epsilon} $, with a value $\epsilon = 1- (2\alpha+\gamma +c) > 0$ that comes from choosing our observable to have a spectral subset of size $\O{\gamma}$ and a resolution $\O{-(1+\alpha)}$. The final contribution comes from viewing the random walk in a time window $\Delta t = \O{c}$. Thus, for large $N$ the use of the continuous Fokker-Planck equation suffices to describe the spectral evolution with a resolution better than the mean level spacing. In subsequent sections we obtain the precise form of this Fokker-Planck equation for three different examples of Bernoulli ensembles, achieved through a careful analysis of the moments of the transition probability. The resulting equations are exactly those obtained by Dyson \cite{Dyson-1962} - thus showing the stochastic motion of the eigenvalues in our Bernoulli ensembles tends to the familiar motion of eigenvalues in the canonical Gaussian ensemble setting.

The first example, in Section \ref{Real symmetric}, concerns real-symmetric Bernoulli matrices. The second (Section \ref{Imaginary anti-symmetric}), concerns imaginary anti-symmetric Bernoulli matrices and the third (Section \ref{Real Wishart}) investigates the singular values of real non-Hermitian $N \times M$ Bernoulli matrices. These ensembles are closely related to the graph types (\ref{Undirected graphs}), (\ref{Tournament graphs}) and (\ref{Directed graphs}) respectively (see above). In the three cases we recover the JPDF corresponding to the fixed trace GOE, antisymmetric GUE (see \cite{Mehtabook,Zhoua-2010}), and real Wishart distributions (see e.g. \cite{Akemann-2011} and references therein) respectively. Finally, in Section \ref{conclusions}, we discuss the results and outline potential developments.

\section{Random walks on Bernoulli matrices}\label{random walks}
Let $\mathcal{B}_N$ denote one of the unconstrained ensembles of $N\times N$ balanced matrices, as defined in the preceding section. Each matrix $B\in \mathcal{B}_N$ consists of entries $B_{ij} \in \{\pm 1\}$, of which $d_N$ of these variables are independent. Hence there are $|\mathcal{B}_N| = 2^{d_N}$ matrices in the ensemble, with each one possessing equal probability to be selected. However, rather than simply selecting a matrix at random we employ an alternative method that allows us to recover this uniform distribution. The idea is to initiate a discrete random walk process over $\mathcal{B}_N$. If this random walk is ergodic then a uniform distribution is recovered in the infinite time limit as the system approaches equilibrium. This walk is a discrete analogue of Dyson's Brownian motion model for the Gaussian matrix ensembles.

To start, let us suppose that each matrix represents a vertex in a large \emph{meta-graph} (so called because it encompasses all graphs in our ensemble). Notice that each matrix is in one-to-one correspondence with the $d_N$-dimensional vector $v(B) = (v_1,\ldots,v_{d_N})$, $v_i \in \{\pm 1\}$. Therefore the vertices of the meta-graph correspond to the corners of a $d_N$-dimensional hypercube with $2^{d_N}$ vertices. Two vertices are then said to be connected, or adjacent (denoted $B \sim B'$), if they differ by exactly one matrix element, or alternatively if their Hamming distance $D(B,B') := \frac{1}{2}||v(B) - v(B')||_1$ is equal to 1. This connectivity relation thus means each matrix/vertex has $d_N$ immediate neighbours.

The analysis of random walks on the hypercube are well established and we refer the reader to \cite{Diaconis,Letac}, and references therein, for more details. Here, following \cite{Diaconis}, we define the transition probability that the walker either stays in the same place, or moves to any neighbour in one time step as equal to $1/(d_N+1)$. Thus, if $P_t(B)$ is the probability to be at $B$ at time $t$, then the probability distribution a single time step later is given by
\begin{equation}\label{hypercube transition}
P_{t+1}(B) = [\varrho P_t](B) = \frac{1}{d_N+1}\sum_{B' : D(B,B') \leq 1} P_t(B').
\end{equation}
This particular Markovian evolution  ensures there is a unique stationary distribution given by $P_{\infty}(B) = 2^{-d_N}$ for all $B$\footnote{In contrast, if one does not allow the random walker the chance to remain in the same place then there is no unique stationary distribution, since the hypercube is bipartite.}. The authors of \cite{Diaconis} perform a comprehensive analysis of this approach to equilibrium by starting with a probability distribution $P_0(o)=1$ localised at the origin, which we choose to be $v(o) = (-1,\ldots,-1)$. They obtain an asymptotic expression for the total variation distance
\begin{equation}\label{Total variation distance}
\mathcal{V}(t) = \sum_{B \in \mathcal{B}_N :P_t(B)\geq 2^{-d_N}} |P_t(B) - 2^{-d_N}| \sim \Erf(e^{-2c(t)}/\sqrt{8})
\end{equation}
in the limit of large $d_N$. Here $\Erf(x) = \int_0^x dz \; e^{-z^2}/(2\pi)$ is the usual error function and $c(t) = (t - t_{\crit})/d_N$. It shows that $\mathcal{V}(t)$ remains very close to 1 until a critical time, calculated to be $t_{\crit}= \frac {1}{4} d_N\log d_N$, after which the probability density relaxes to the equilibrium distribution exponentially fast in the region $t > t_\crit$ and large $d_N$ limit.
\begin{equation}\label{Asymptotic variation distance}
\mathcal{V}(t) \sim \frac{1}{ \sqrt { 2\pi}}\exp\left(-2\frac{t-t_{\crit}} {d_N}\right).
\end{equation}
This exponential approach to equilibrium as $t \to \infty$ is controlled by the second largest eigenvalue (the largest taking the value 1) of the transition probability matrix $\varrho$ in (\ref{hypercube transition}). Due to the symmetry of the hypercube, the eigenvalues of $\varrho$ are equivalent, up to degeneracies, with those of the transition probability $\rho$, which describes the distance from the origin, outlined in Section \ref{entropic forces}. These are given by $\Lambda_j = (1 - \frac{2j}{d_N + 1})$ for $j=0,\ldots d_N$. Hence, we find the associated quantity $C(t) = ||\varrho^t\delta_o - P_{\infty}|| \sim (1 - \frac{2}{d_N + 1})^t$ (where $\delta_o$ denotes the probability density that is equal to 1 at the origin and 0 everywhere else) displays the same asymptotic approach to equilibrium (\ref{Asymptotic variation distance}). In particular one observes this occurs at a rate controlled by the hypercube dimension $d_N$ and so we define a rescaled time $\eta = t/d_N$, which is the natural time-scale in this context.

\section{Induced random walks and entropic forces}\label{entropic forces}
In Section \ref{Spectral statistics} we analyse the joint probability distribution for the eigenvalues by studying how the random walk in $\mathcal{B}_N$ induces a corresponding random walk in the space of spectra. This projection amounts to a reduction of the information about the matrix. As a result, points in our ensemble, that were once distinct, become indistinguishable, or much closer, in the reduced space. This change of viewpoint creates effective forces that originate due to entropic, rather than mechanical, reasons. Consequently, one obtains a Fokker-Planck equation that describes how the random walker exhibits preferential concentration in certain regions in the space of spectra.

\begin{figure}[ht]
\includegraphics[width=0.85\textwidth]{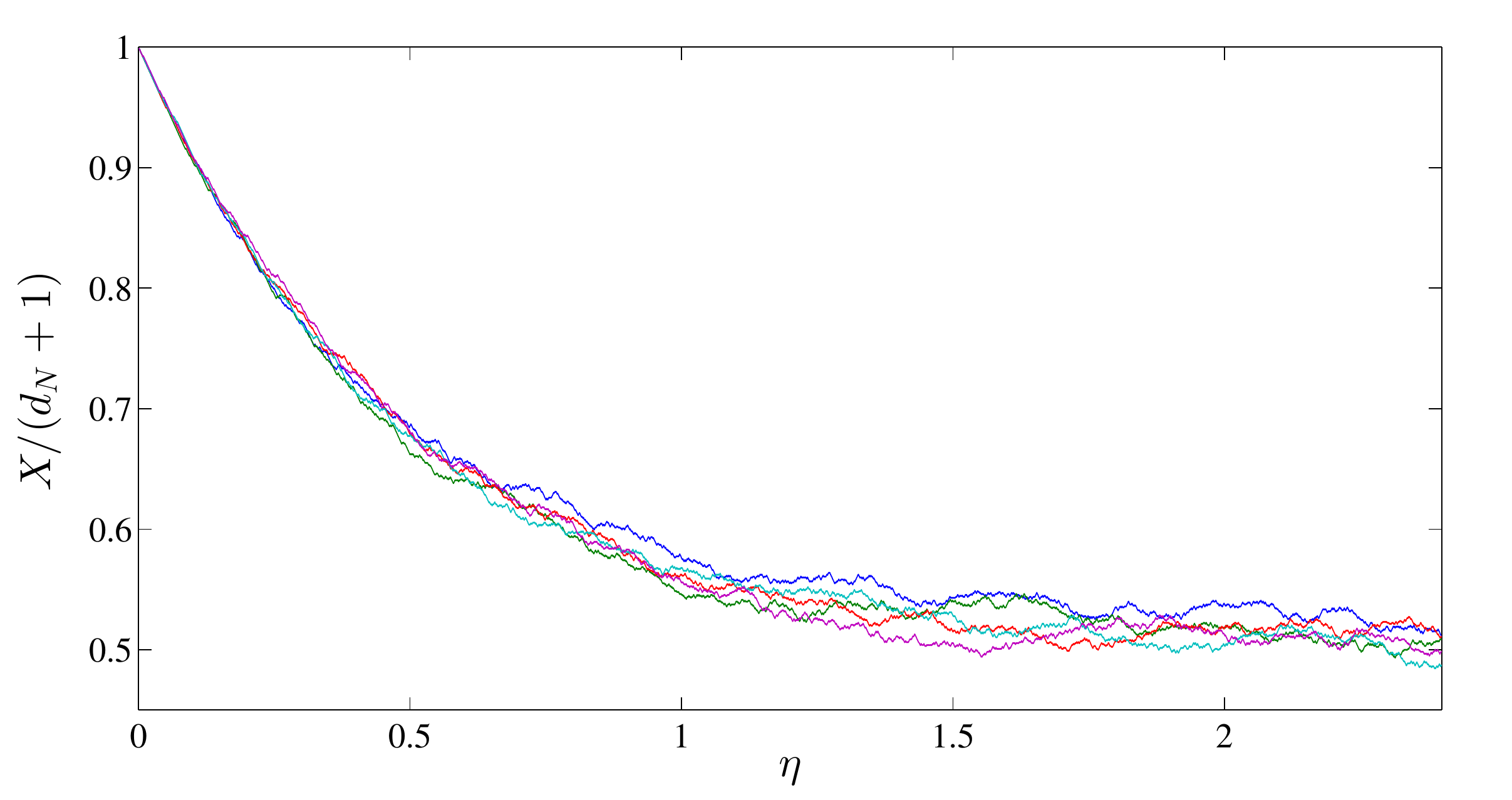}
\caption{Five simulated random walks over the space of matrices (\ref{Real Bernoulli}), with $N=50$ and an initial matrix $B^{(0)}$ satisfying $B^{(0)}_{ij}=1 \; \forall \; i,j$. Plotted is $X(\eta)/(d_N+1)$ as a function of $\eta = t/d_N$, with saturation to equilibrium occurring at approximately $\eta_{\crit} = t_{\crit}/d_N = \log(d_N)/4 \approx 1.78$.}
\label{fig:X(t)}
\end{figure}

Before we progress, as a prelude, we provide a brief example that illustrates this phenomenon: We compute the distribution of the Hamming distance of the random walker from the origin $X(B) = D(B,o)$. This is in fact equal to $\Tr AA^\dag = \sum_k \lambda_k^2(A)$ (where $A$ is the adjacency matrix). It is therefore the simplest instance in which the random walk relates directly to the spectrum. A numerical simulation of this walk is given in Fig. \ref{fig:X(t)}, where several random trajectories of $X(B(t))$, all starting at the origin $o$, reach equilibrium around some critical time $t_{\crit}$. Similar problems, in various other contexts, are well discussed in a number of articles \cite{Wang,Kac,Letac,Diaconis}, to which we refer the reader for more details. Here we shall outline the essential points.

To begin, notice that $X(B)$ is given by the number of 1's in the code $v(B)$, therefore the probability that the walker moves from $B$ to another $B'$ with $X(B')=X(B)-1$ is proportional to $X$. Similarly the likelihood of increasing $X$ by 1 is proportional to $d_N-X$. Given this transition probability is the same for all such matrices we drop the explicit reference to $B$ and write
\begin{equation}\label{time evolution}
\rho_{X\to X-1} =\frac{X}{d_N+1}, \hspace{10pt}
\rho_{X\to X+1} = \frac{d_N-X}{d_N+1}, \hspace{10pt}
\rho_{X\to X} = \frac{1}{d_N+1},
\end{equation}
with 0 otherwise. Therefore, if $P_t(X)$ is the probability to find a vertex with a given Hamming distance $X$ at time $t$,  then the probability distribution is updated at the next time step via the following process
\begin{eqnarray}\label{eq:smul}
\hspace{-25mm}  P_{t+1}(X)-P_t(X)
&=&P_t(X-1)\rho_{X-1\rightarrow X}+P_t(X+1)\rho_{X+1\rightarrow X}-P_t(X)(1 - \rho_{X\to X}) \nonumber \\
&=& \frac{d_N}{2(d_N+1)} \left(P_t(X+1)+P_t(X-1)-2P_t(X)\right )  \nonumber \\
&+&  \left(  P_t(X+1)\frac{X+1-d_N/2}{d_N+1} - P_t(X-1)\frac{X-1-d_N/2}{d_N+1}\right ) .
\end{eqnarray}
A detailed study of this discrete Markovian evolution can be found in \cite{Kac,Letac}. The spectrum of the time evolution operator (\ref{time evolution}) is given by $\Lambda_j = (1-\frac{2j}{d_N+1})$, with $j=0,1,\ldots,d_N$. As mentioned previously, this spectrum coincides with the evolution operator for the uniform hypercube in Section \ref{random walks}. The authors of \cite{Diaconis} therefore utilise the solution of (\ref{eq:smul}) in order to analyse the approach to equilibrium of (\ref{hypercube transition}). The stationary distribution for $P_t(X)$ corresponds to the first eigenvalue of $\rho$ (given by $\Lambda_0 = 1$) and is reached when the cube is uniformly covered:
\begin{equation}\label{Stat distribution Gaussian}
\hspace{-20pt}P_{\infty}(X) = \frac{1}{2^{d_N}} \left (
\begin{array} {c}
d_N\\
X
\end{array}
\right ) \to {\sqrt \frac{2}{\pi d_N}} \exp \left [-\frac{2}{d_N}(X-\frac{d_N}{2})^2 \right], \hspace{5pt} N \to \infty .
\end{equation}
The asymptotic estimate above is only valid for $|X-\frac{d_N}{2}|\in {\it o}(d_N^{\frac{3}{4}})$ , and it gets worse when $X$ is farther away from  $\frac{d_N}{2}$. At the centre, the error is of order $\mathcal{O}(d_N^{-\frac{3}{2}})$. In the continuous limit, as we approach equilibrium the probability density becomes normally distributed with mean $\mu = d_N/2$ and variance $\sigma^2 = d_N/4$. It shows that the majority of our vertices on the hypercube lie in a region of size proportional to $\sqrt{d_N}$ centred at a distance $d_N/2$ away from the origin. Thus our random walker is `attracted' to this region simply because the number of vertices outweigh those near the origin.

The result (\ref{Stat distribution Gaussian}) can also be obtained by taking the continuous limit of the equation (\ref{eq:smul}) instead. This is done by  introducing a scaled {\it spatial} variable $\xi = \frac{X - d_N/2}{\sqrt{d_N}} \in [-\sqrt{d_N}/2,\sqrt{d_N}/2]$ and a scaled  {\it temporal} variable $\eta=t/d_N$.  Then, in the limit   the discrete evolution equation approaches a Fokker-Planck (Smoluchowski) equation describing an Ornstein-Uhlenbeck process \cite{Wang,Kac,Letac}:
\begin{equation}\label{UO-process}
\frac {\partial P(\xi,\eta )}{ \partial \eta} = \frac{1}{2}\frac {\partial^2 P(\xi,\eta )}{ \partial \xi^2} + 2\frac {\partial(\xi P(\xi,\eta ))}{\partial \xi}.
\end{equation}
Setting the LHS equal to 0 and solving for $P(\xi,\infty )$, one obtains (\ref{Stat distribution Gaussian}) in its scaled form. Several comments are in order at this point:

\begin{enumerate}
\item Choosing a different scaling of $X$, such as $x=X/d_N-1/2$ with $x\in [-1/2,1/2]$ might have appeared natural in some instances, and would lead to $P(x,\infty )=\delta(x)$, emphasising the concentration of the probability to the centre of the $X$ interval. However, one then loses the important information that in the limit, the variable $\xi$ behaves as a Brownian process.

\item \label{Scaling relations} The scaling of $X$ with $\sqrt{d_N}$ is the natural scaling in the present context. It is equivalent to scaling $\Tr AA^\dag$ by $N$ and hence $A$ by $\sqrt{N}$. For this reason we employ the same scaling when investigating the spectra of $\mathcal{B}_N$ in Section \ref{Spectral statistics}. Similarly, the time scaling $\eta = t/d_N$ will also be used in Section \ref{Spectral statistics}.

\item The second term in the RHS of the Fokker-Planck (\ref{UO-process}) is usually interpreted as  due to an  effective drift force, which takes here the form  $F(\xi) = -2\xi$  as if it  derives from a confining harmonic potential. Clearly, in the present context it arises purely because of the combinatorial (entropic) preferences in the hypercube, which create the illusion of a confining harmonic potential. This mean drift towards the centre $\xi=0$ will also appear when we investigate the stochastic nature of the eigenvalues with respect to perturbations of the matrix $B$. Note also that equation (\ref{UO-process}) plays a crucial r\^ole in Dyson's Coulomb gas model for random matrices \cite{Dyson-1962}, which he uses in order to introduce the assumed Brownian motion of matrix elements for the Gaussian ensembles.
\end{enumerate}

\section{Spectral statistics}\label{Spectral statistics}
We now turn our attention to the induced random walk in the space of spectra. The formalism will be presented in general and then applied systematically to specific ensembles of Bernoulli matrices in Sections \ref{Real symmetric}, \ref{Imaginary anti-symmetric} and \ref{Real Wishart}. Following from the discussion of point (\ref{Scaling relations})) in the previous section we introduce the scaled matrix $\bar{B} = B/\sqrt{N}$, to which we associate an $N$-dimensional vector of eigenvalues (or singular values as will be the case in Section \ref{Real Wishart})
\begin{equation}\label{Spectral vector}
\bsigma(\bar{B}) = (\lambda_1(\bar{B}) \leq \lambda_2(\bar{B}) \leq \ldots \leq \lambda_N(\bar{B})).
\end{equation}
This ensures the eigenvalues $\lambda_{\nu}$ are $\mathcal{O}(1)$ (and thus the mean level spacing is $\mathcal{O}(N^{-1})$). Furthermore, in each ensemble, we shall find that every matrix $\bar{B} \in \mathcal{B}_N$ satisfies $\Tr(\bar{B}^\dagger \bar{B}) = R_N^2$, where $R_N$ is an $N$-dependent constant (to be calculated) of size $\mathcal{O}(N^{1/2})$. Hence, the set of spectra, which we denote by $\Sigma_N$, forms a \emph{discrete} and \emph{finite} subset of the continuous space $S_N = \{ \bs \in \mathbb{R}^N : s_{\nu} \leq s_{\nu+1}, || \bs ||_2 = R_N\}$.

\begin{figure}[ht]
\includegraphics[width=0.9\textwidth]{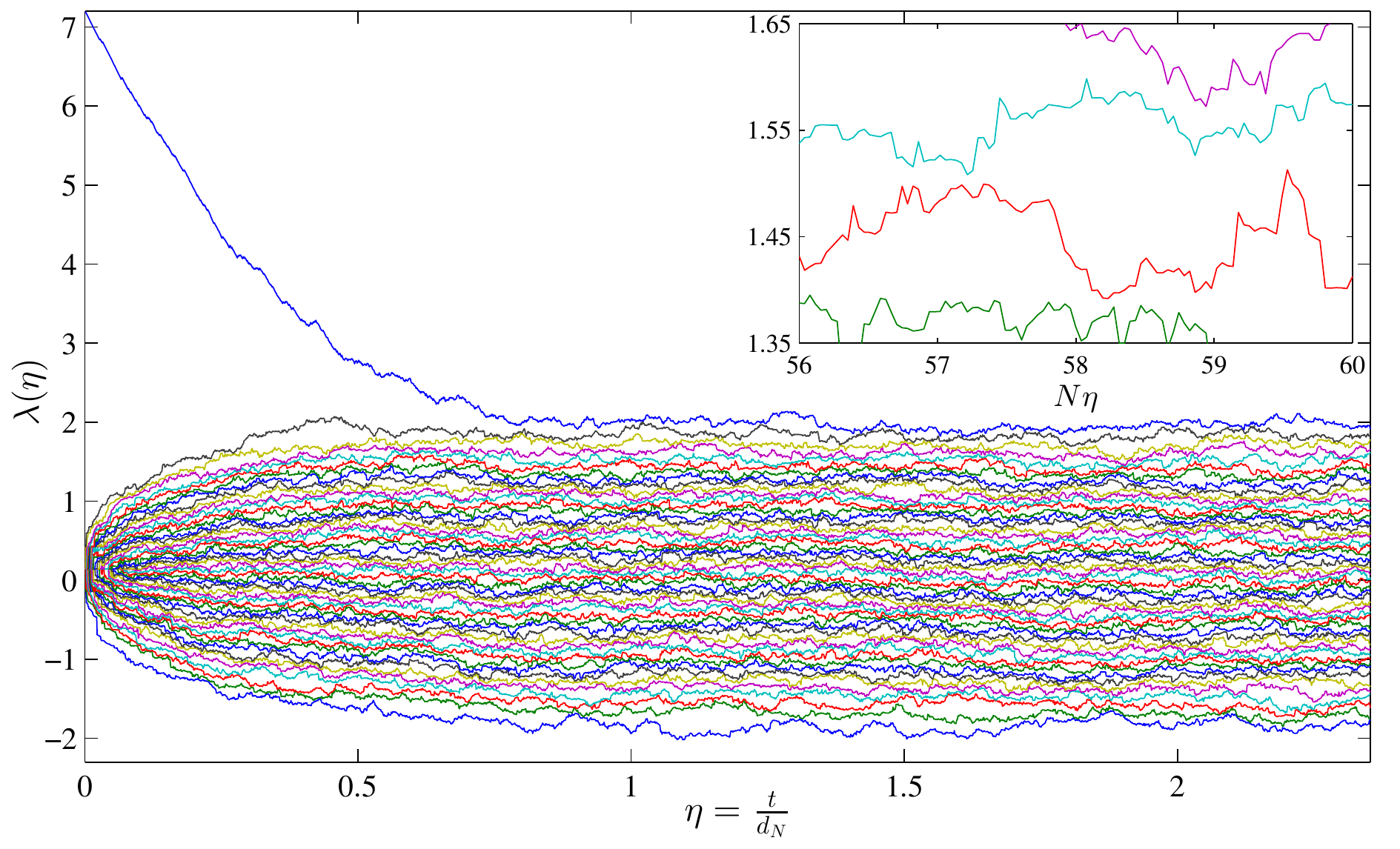}
\caption{A single instance of the spectral evolution during the same random walk process as Fig. \ref{fig:X(t)}. Plotted is the entire spectrum as a function of $\eta$. The inset provides a close up view of the same process.}
\label{fig:sigma(t)}
\end{figure}

Suppose now we initiate our random walk, starting at $\bar{B}^{(t=0)}$ and moving through the meta-graph according to the transition probability (\ref{hypercube transition}). What are the dynamics of $\bsigma(\bar{B}^{(t)}) \in \Sigma_N$ as we follow this walk over the integer-valued time $t$? Fig. \ref{fig:sigma(t)} provides a numerical simulation of a `typical' realisation of this walk: Every eigenvalue in $\bsigma$ is plotted as a function of the scaled time $\eta =  t/d_N$. The main point to observe here is the clear separation between the microscopic, mesoscopic and macroscopic times-scales. In the main figure we see how the eigenvalues approach equilibrium on scales of order of the system size $t = \mathcal{O}(d_N)$ (in contrast to Fig. \ref{fig:X(t)} in which it was $d_N\log(d_N)/4$), however before reaching this stage coherent trajectories already emerge due to the drift forces associated with the `repulsion' of eigenvalues. This happens on a time scale $t = \mathcal{O}(N)$, which is more evident in the inset picture. In the microscopic regime we have fluctuations arising from a change in the spectrum $\delta \bsigma$ due to a single time step, which we denote $\delta t = 1$. Between these sits the mesoscopic regime which characterises the point at which these spectral trajectories start to emerge. Here we obtain fluctuations $\Delta \bsigma$ corresponding to small (but still larger than microscopic) changes in time $\Delta t = \O{c}$, $0<c<1$. In the remainder of the paper we shall continue with this convention of using $\delta$ to denote variations due to a single step and $\Delta$ to denote variations due to $\Delta t$ time steps. The clear hierarchy of time-scales is one of the key ingredients which enables the construction of a Fokker-Planck approach to describe the spectral evolution which emerges due to the underlying random walk.

The evolution of the spectral distribution on the microscopic time scale starts with a Markovian master equation analogous  to (\ref{eq:smul}):
Let  $\pi(\bsigma,t)$,  be the  probability  to be  at some point $\bsigma\in\Sigma_N$ at time $t$, subject to some prescribed initial conditions.\footnote{Note that at this level of the description, both the time $t$ and the space coordinates $\bsigma$ are discrete variables.} Then for $b(\bsigma) = \{\bar{B} \in \mathcal{B}_N : \bsigma(\bar{B}) = \bsigma \}$ we have
\begin{equation}
\label{micr.evol.}
\pi(\bsigma,t+1) = \frac{1}{|b(\bsigma)|}\sum_{\bar{B} \in b(\bsigma)} \frac{1}{d_N+1}\sum_{\bar{B}': \bar{D}(\bar{B},\bar{B}') \leq 1}\pi(\bsigma(\bar{B}'),t),
\end{equation}
where $\bar{D}(\bar{B},\bar{B}') = D(B,B')$. In the limit $t \to \infty$, $\pi(\bsigma,t)$ will converge to the stationary distribution associated to the JPDF for the eigenvalues of $\mathcal{B}_N$. However, in contrast to the example in Section \ref{entropic forces}, there is no closed form expression for $\pi(\bsigma,t)$, or even $\pi(\bsigma,\infty)$. Instead, the best we can hope for is a convergence as $N \to \infty$ to a limiting distribution, analogous to the Ornstein-Uhlenbeck process (\ref{UO-process}) in the previous section. Unfortunately, the obvious relationship between (\ref{UO-process}) and its discrete process (\ref{eq:smul}) has no analogue in the context of the induced random walk in the space of spectra. For this reason, we introduce an alternative derivation that overcomes these failures. It is an adaptation of Kolmogorov's method \cite{Kolmogorov}, which is eloquently presented in Section 8 (c) of Wang and Uhlenbeck \cite{Wang} (other approaches using essentially the same arguments are also outlined in \cite{Risken-1989,Gardiner-2009}).

In the classic example of a random walker on a 1D line (see e.g. \cite{Gardiner-2009} Section 3.8.2), the Brownian motion formalism emerges from its underlying microscopic dynamics because the motion of the walker is described exactly by a sum of independent random variables. In other words the central limit theorem (CLT) guarantees for a time window of $\Delta t$, the transition probability is Gaussian, up to an error $(\Delta t)^{-1/2}$ due to the Berry-Esseen theorem (see for example \cite{Tao-book,Tao-2011b} for various versions of this theorem and previous applications to RMT). Whilst we shall not invoke the CLT explicitly in our context, it is helpful to keep this line of thought in mind, as it underlies the transition from the microscopic discrete view provided  by (\ref {micr.evol.}) to a coarse grained Fokker-Planck description. This is made possible due to the clear separation between the microscopic and macroscopic time scales. On the intermediate (mesoscopic) time scale $\Delta t \sim N^{c}, \ 0<c<1$, our evolution will be given as an accumulation of, approximately, independent random steps. At the same time, the spatial resolution is deteriorated due to the increasing width of the distribution acquired during  $(\Delta t)$, which will be shown to be of order $N^{-\frac{3-c}{2}}$. Thus, the temporal and spatial coarse graining have opposite dependence on $\Delta t$, and an optimum description should be sought by the freedom of choice of the parameter $c$. The embedding of the discrete evolution in continuous space-time is made possible by the fact that the macroscopic time scale is $\mathcal{O}(d_N) = \mathcal{O}(N^2)$, so that on this scale $\Delta t$ can be approximated by a differential. However, one should bear in mind that as long as $N$ is finite, the continuous space-time evolution is not precise and the errors accumulated by various approximations will be discussed in detail.

\subsection{The Fokker-Planck equation}\label{fokker-planck}

Evolving our random walk by a single time-step equates to choosing an entry $\bar{B}_{pq}$ at random from the matrix $\bar{B}$ and changing its sign. This produces a new matrix $\bar{B}'$, with the difference matrix $\delta \bar{B} = \bar{B}' - \bar{B}$ of either rank 1 or 2. The form of $\delta \bar{B}$ with respect to this change is given, for example, in (\ref{Matrix perturbation}). In $\Delta t$ time steps the random walker is capable of moving from $\bar{B}$ to another matrix $\bar{B}'$, a maximum Hamming distance $\bar{D}(\bar{B},\bar{B}') = \Delta t$ away. This corresponds to a path $\omega = (\omega_1,\ldots,\omega_{\Delta t})$ in our hypercube $\mathcal{B}_N$, where each $\omega_r$ denotes the swapping of a particular matrix element $pq$.\footnote{There is also the possibility of remaining in the same place, which is treated more thoroughly in Section \ref{moments}.} The perturbation associated with $\omega$ is simply the addition of all single step perturbations along this path
\begin{equation}\label{Perturbation - multi-step}
\Delta \bar{B}^{\omega} = \bar{B}' - \bar{B} = \sum_{r=1}^{\Delta t} \delta \bar{B}^{\omega_r}.
\end{equation}

The probability to move from the matrix $\bar{B}$ to $\bar{B}' = \bar{B} + \Delta \bar{B}$ in $\Delta t$ time steps is denoted $\rho_{\Delta t}(\bar{B} \to \bar{B}')$ and given by $\varrho^{\Delta t}$, as defined in (\ref{hypercube transition}). Employing this coarse-graining in time the microscopic Master equation (\ref {micr.evol.}) becomes (abusing notation slightly and writing in terms of the rescaled time variable $\eta$)
\begin{equation}
\label{coars.evol.}
 \pi(\bsigma,\eta + \Delta \eta) =  \sum_{\bsigma' \in \Sigma_N} \rho_{\Delta \eta}(\bsigma' \to \bsigma) \pi(\bsigma',\eta).
\end{equation}
Here the transition probability in the space of spectra $\Sigma_N$ is given in terms of the underlying evolution in the space of matrices, i.e.
\begin{equation}
\fl
\rho_{\Delta \eta}(\bsigma\rightarrow\bsigma')  := \frac{1}{|b(\bsigma)|}\sum_{\bar{B} \in b(\bsigma)} \sum_{\bar{B}' : \bar{D}(\bar{B},\bar{B}') \leq \Delta t } \rho_{\Delta t}(\bar{B} \to \bar{B}') \; \delta(\bsigma(\bar{B}'),\bsigma'),
\end{equation}
where $\delta(a,b)$ is the Kronecker delta. However, in Section \ref{moments} we show (in the large $N$ limit) this transition is independent of the initial matrix $\bar{B}$ and so rather than summing over all matrices in $b(\bsigma)$ we can choose any particular matrix $\bar{B}$ that satisfies $\bsigma = \bsigma(\bar{B})$.
\begin{equation}\label{Transition probability}
\rho_{\Delta \eta}(\bsigma\rightarrow\bsigma')  := \sum_{B' :\bar{D}(\bar{B},\bar{B}') \leq \Delta t } \rho_{\Delta t}(\bar{B} \to \bar{B}') \; \delta(\bsigma(\bar{B}'),\bsigma'),
\end{equation}

The next step in the derivation requires the construction of the time-averaged moments of the evolution,
\begin{equation}\label{moments definition}
\fl
M_{\nu_1\ldots\nu_k} := \frac{\E(\prod_{i=1}^k\Delta \lambda_{\nu_i})}{\Delta \eta}\  = \frac{1}{\Delta \eta} \sum_{\sigma' \in \Sigma_N} \prod_{i=1}^k(\lambda'_{\nu_i} - \lambda_{\nu_i})\rho_{\Delta \eta}(\bsigma\rightarrow\bsigma')
\end{equation}
from (\ref{Transition probability}).

If we had a continuous process, as in Dyson's prescription of the Gaussian ensembles \cite{Dyson-1962}, the derivation would proceed by showing that only the first two moments survive the limit  $\Delta \eta \to 0$ \cite{Wang}. In the present case our process is discrete and therefore for any finite $N$ there is bound on how small we can take $\Delta \eta$. For this reason, one can only show that for some finite $N$ the first two moments are (to leading order) independent of $\Delta \eta$ and that higher moments will depend on higher powers of $\Delta \eta$. This persistent presence of the higher terms is the main source of error in the transition from the discrete to the continuous description.

\vspace{10pt}

Following the approach in \cite{Wang}, we introduce a test function $h(\bsigma)$, which vanishes on the boundary $S_N$, together with its derivative. The requirements on the smoothness of $h(\bsigma)$ and on its domain will be specified later. We now study the coarse grained time evolution of the expectation value
\begin{equation}
\label{exph}
h_{\eta} :=\sum_{\bsigma \in \Sigma_N} h(\bsigma) \pi(\bsigma,\eta ).
\end{equation}
Following (\ref{coars.evol.}) we have
\[ h_{\eta+ \Delta \eta}=
\sum_{\bsigma \in \Sigma_N} h(\bsigma) \pi(\bsigma,\eta + \Delta \eta) =  \sum_{\bsigma,\bsigma' \in \Sigma_N} \rho_{\Delta \eta}(\bsigma \to \bsigma') \pi(\bsigma,\eta)h(\bsigma').
\]
Expanding $h(\bsigma)$ as a Taylor series, dividing through by $\Delta \eta$ and performing the sum over $\bsigma'$ we obtain, using (\ref{moments definition}),
\begin{equation}\label{FP expansion new}
\fl
\sum_{\bsigma \in \Sigma_N} h(\bsigma) \frac{\Delta\pi}{\Delta \eta} =  \sum_{\bsigma \in \Sigma_N}\pi(\bsigma,\eta)\left[ \sum_{\nu} M_{\nu}\frac{\partial h}{\partial \lambda_{\nu}} + \sum_{\nu\mu} \frac{M_{\nu\mu}}{2} \frac{\partial^2 h}{\partial \lambda_{\nu}\partial \lambda_{\mu}} + \ldots
\right]
\end{equation}
where $\Delta \pi = \pi(\bsigma,\eta + \Delta \eta) - \pi(\bsigma,\eta)$.

It was shown above that the coarse grained temporal description induces naturally a corresponding spatial coarse grained representation of $\pi(\bsigma,\eta)$, and therefore it is advantageous to convert the sum over the discrete set of spectral points $\Sigma_N$ to an integral over the covering domain $S_N$. In contrast to the previous section, in which the points $X$ were arranged in a lattice, the spectra $\bsigma$ are distributed through $S_N$ in some random fashion. We therefore assume the existence of a density function $q(\bsigma)$, so the set $\Sigma_N$ effectively appears to be a sample of $|\Sigma_N|$ points chosen randomly from the distribution $q(\bsigma)$. Given this assumption the expectation of some function $f$ with respect to $q(\bsigma)$ is
\[
\int_{S_N} d\bsigma \; f(\bsigma) q (\bsigma)  \approx \frac{1}{|\Sigma_N|}\sum_{\sigma \in \Sigma_N} f(\bsigma) \equiv \E_{\Sigma_N}(f).
\]
This approximation is the premise of the Monte-Carlo integration technique. If indeed the points $\bsigma$ can be assumed to be random distributed points according to $q(\bsigma)$ then the error is essentially given by the law of large numbers, i.e. $\sqrt{\Var(f)/|\Sigma_N|}$, where
\[
\Var(f) = \frac{1}{|\Sigma_N|}\sum_{\bsigma \in \Sigma_N} (f(\bsigma) - \E_{\Sigma_N}(f))^2.
\]
Importantly, this error does not depend on the dimension of the space, only on the number of points $|\Sigma_N|$. Asymptotically this is $|\Sigma_N| \sim 2^{d_N}/N!$, which grows exponentially as $N$ increases, meaning the error decreases very fast. Moreover, the density of spectra is
\[
\frac{|\Sigma_N|}{|S_N|} \sim \frac{2^{N-1}\Gamma(N/2)}{2\pi^{N/2}(N+1)} \frac{2^{N(N+1)/2}}{\Gamma(N+1)} \sim e^{N^2\ln 2 - z N\ln N}, \hspace{5pt} z >0.
\]
This leads to an exponential increase in the number of spectra per unit (hyper) volume, allowing us to take functions $f(\bsigma)$ on a scale far smaller than the typical spacing between eigenvalues. We therefore write
\[
\int_{S_N} d\bsigma \; f(\bsigma) \bar{\pi}(\bsigma,\eta)q(\bsigma) \approx \sum_{\bsigma \in \Sigma_N}\pi(\bsigma,\eta) f(\bsigma),
\]
with $\bar{\pi}(\bsigma,\eta)$ a smooth approximation of $|\Sigma_N|\pi(\bsigma,\eta)$. In fact the main source of error will come from the higher terms in the expansion (\ref{FP expansion new}) and this sets a much greater restriction on our test functions. Writing $Q(\bsigma,\eta) = \bar{\pi}(\bsigma,\eta)q(\bsigma)$  and replacing the sum in (\ref{FP expansion new}) by an integral we arrive at the following expression
\begin{equation}\label{Pre-FP equation}
\fl
\int_{S_N} d\bsigma \; h(\bsigma)\frac{\Delta Q}{\Delta \eta} = \int_{S_N} d\bsigma \;Q(\bsigma,\eta )\left[
\sum_{\nu} M_{\nu}\frac{\partial h}{\partial \lambda_{\nu}} +\frac{1}{2} \sum_{\nu
\mu} M_{\nu\mu}  \frac{\partial^2 h}{\partial \lambda_{\nu}\partial \lambda_{\mu}}\right] + \mathcal{E}(N,h(\bsigma)).
\end{equation}
The error $\mathcal{E}(N,h(\bsigma))$ depends on a number of factors and will be discussed below in detail. If we replace the time difference in the LHS side of (\ref{Pre-FP equation}) by a derivative (by sacrificing an error of $\mathcal{O}(\Delta \eta) = \O{c-2}$), and  integrate by parts  the RHS of (\ref{Pre-FP equation}) (where we make use of the requirement that $h$ and its gradient vanish on the boundary of $S_N$) we get
\begin{equation}\label{FP eqn main}
\fl
\int_{S_N} d\bsigma \; h(\bsigma) \left\{ \frac{\partial Q}{\partial \eta} +
 \sum_{\nu} \frac{\partial( M_{\nu} Q)}{\partial \lambda_{\nu}} - \frac{1}{2}\sum_{\nu \mu}\frac{\partial^2(M_{\nu\mu} Q)}{\partial \lambda_{\nu}\partial\lambda_{\mu}}\right\}  = \mathcal{E}(N,h(\bsigma)).
\end{equation}
Consider the above relation where the error term is replaced by $0$. In order to satisfy this equation independently  of the actual choice of the observable $h$, the expression in the curly brackets must vanish. Thus $Q(\bsigma,\eta)$ should  satisfied the Fokker-Planck equation:
 \begin{equation}\label{FP eqn finally}
 \frac{\partial Q}{\partial \eta} =
 -\sum_{\nu} \frac{\partial( M_{\nu} Q)}{\partial \lambda_{\nu}} + \frac{1}{2}\sum_{\nu \mu}\frac{\partial^2(M_{\nu\mu} Q)}{\partial \lambda_{\nu}\partial\lambda_{\mu}}.
\end{equation}
However $\mathcal{E}(N,h(\bsigma))$ does not vanish. At best, under certain smoothness constrains on $h$ the error term vanishes as a negative power of $N$ in the limit $N\rightarrow \infty$. Hence, for finite $N$ the evolution of the spectra under the random walk is described by the solution of the Fokker-Planck equation within an error arising from a number of factors, as detailed below. We also note that it will transpire that the off-diagonal second order moments $M_{\nu\mu}$ (with $\nu \neq \mu$) provide a lower order contribution than their diagonal counterparts $M_{\nu\nu}$. Therefore, it is amenable to remove these terms, and place their contribution within the error term $\mathcal{E}(N,h(\bsigma))$. Doing so, we shall obtain a Fokker-Planck equation in Sections  \ref{Real symmetric}, \ref{Imaginary anti-symmetric} and \ref{Real Wishart} with only a single sum over $\nu$, which is consistent with the resulting equation in Dyson's approach \cite{Dyson-1962}.

\subsection{Error estimates}
\vspace{10pt}
There are several sources of error which occur in the transition from the discrete to the continuous description of the spectral evolution for finite $N$:

\begin{enumerate}
\item
{\it The replacement of discrete sums (over space) and differences (in time) by integrals and differentials, respectively.}  The error induced by the former was discussed above and shown to be exponentially small, while the latter is of order $\mathcal{O} (N^{-(2-c)})$ .
\item
{\it Corrections to the leading order expressions for the moments $M_{\nu}$ and $M_{\nu\nu}$}. In all three ensembles we consider, these corrections will arise from higher terms in the perturbation expansion and will require assumptions regarding the delocalisation of eigenvectors and the rigidity of eigenvalues. The conditions under which the assumptions are valid have recently been stated (see e.g. \cite{Tao-2011b,Tao-2012,Rudelson-2014} and references therein for details) and are known to be satisfied  with a high probability (tending to 1 in the limit of large $N$).
 \item
 {\it Off-diagonal second order moments $M_{\nu\mu}$ (with $\nu \neq \mu$)}, as described above.
 \item
 {\it Moments in (\ref {FP expansion new}) of order 3 and higher.}
\end{enumerate}

This last source of error will in fact provide the greatest contribution to $\mathcal{E}(N,h(\bsigma))$ which, as seen in (\ref{FP expansion new}) is coupled to higher derivatives of $h(\bsigma)$. To ensure convergence (i.e. for $\mathcal{E}(N,h(\bsigma))$ to decrease in $N$) we must therefore restrict our observables $h(\bsigma)$ to a certain class with appropriate resolution determined by the following restrictions:

1.  The support of its Fourier transform is restricted to a ball of radius $k=\mathcal{O}(N^{1+\alpha}), \ \alpha>0$, which means the derivatives of $h$, in any direction, are of order $h^{(k)} = \O{k(1+\alpha)}h$. In particular, this choice ensures that the resolution is always better than the typical spacing between eigenvalues, which is $\O{-1}$, as the width of $h$ is of the order of $k^{-1}$.

2. The test function $h(\bsigma)$ is defined over a spectral set $\Gamma = \{\lambda_{\nu_i}\}_{i=1}^n$ which consists of $n$ eigenvalues with $ n = \O{\gamma}, \ \gamma>0$. This limits the size of the spectral interval under study to be smaller than the entire spectral support but to include a large number of eigenvalues.

Under the restrictions above, it is possible to estimate the magnitude of the neglected terms. For the purposes of clarity we defer the computations to Sections \ref{moments} and \ref{Higher moments}. Here, we simply mention they are consistent with the results one would suspect from using the CLT in combination with perturbation theory. Namely, that all moments beyond first order are consistent with a Gaussian kernel $\rho_{\Delta\eta}(\bsigma, \bsigma')$ comprising independent $\Delta \lambda_{\nu}$, each with mean $\E(\Delta \lambda_{\nu}) \approx \E(\langle \nu | \delta B |\nu\rangle)\Delta t = \mathcal{O}(1)\Delta \eta$ and variance $\E(\Delta \lambda^2) \approx \E(\langle \nu | \delta \bar{B} |\nu\rangle^2)\Delta t = \O{-1}\Delta \eta$ (the exact expressions are given in Section \ref{moments}). Only the first moment will deviate from this prescription due to the influence of the second term in the perturbation formula (\ref{Perturbation formula}), though it will remain of the same order. Therefore, summarising the contributions we have.

\begin{itemize}
\item \emph{First order}: The first moment is of order $\mathcal{O}(1)$, so
\[
\int d\bsigma Q \sum_{\nu \in \Gamma} M_{\nu} \frac{\partial h}{\partial \lambda_{\nu}} = \int d\bsigma h Q [ \O{1+\gamma + \alpha}],
\]
using that $h' = kh = \O{1+\alpha}h$
\item \emph{Second order:} The main contribution to second order comes from the diagonal moments $M_{\mu\nu} = \O{-1}$. Therefore
\[
\int d\bsigma Q \sum_{\nu \in \Gamma} M_{\nu\nu} \frac{\partial^2 h}{\partial \lambda_{\nu}^2}  = \int d\bsigma h Q [\O{2 + 2\alpha + \gamma -1}],
\]
using that $h'' = k^2h = \O{2+2\alpha}h$. Note the off-diagonal terms will give an overall contribution of
\[
\int d\bsigma Q \sum_{\nu\neq\mu \in \Gamma} M_{\nu\mu} \frac{\partial^2 h}{\partial \lambda_{\nu}\partial \lambda_{\mu}}  = \int d\bsigma h Q [\O{2 + 2\alpha + 2\gamma + c -3}],
\]
which is much less than the diagonal terms.
\item \emph{Third order:} The main third order term comes from moments of the form $M_{\nu\nu\mu}$. Therefore we take
\[
\int d\bsigma Q \sum_{\nu,\mu \in \Gamma} M_{\nu\nu\mu} \frac{\partial^3 h}{\partial \lambda_{\nu}^2\partial \lambda_{\mu}} =
\int d\bsigma h Q [\O{c-3+3+3\alpha+2\gamma}]
\]
\item \emph{Fourth order:} The fourth order term comes from moments of the form $M_{\nu\nu\mu\mu}$
\[
\int d\bsigma Q \sum_{\nu,\mu \in \Gamma} M_{\nu\nu\mu\mu} \frac{\partial^4 h}{\partial \lambda_{\nu}^2\partial \lambda_{\mu}^2} =
\int d\bsigma h Q [\O{c-4+4+4\alpha+2\gamma}]
\]
\end{itemize}
We require the third order to be less than the second. Hence $1 + 2\alpha + \gamma > c +3\alpha + 2\gamma$, or $\alpha + \gamma + c <1$. Similarly, for fourth order we get $1 + 2\alpha + \gamma > c + 4\alpha + 2\gamma$, or $2\alpha + \gamma + c <1$. The latter provides a slightly tighter restriction on $\alpha$. Thus, if   $2\alpha+\gamma +c = 1-\epsilon$, the error estimated for $h_{\eta}$  (\ref {exph}) decreases as $N^{-\epsilon}$.

Ideally, we would like to translate the result (\ref{FP eqn main}) into a statement comparing the JPDF for our Bernoulli ensembles, with the resulting stationary distribution $Q(\bsigma,\infty)$ of the Fokker-Planck equation (\ref{FP eqn finally}). For instance, a weak convergence estimate given by the difference in the expectation of our observable. At present we do not know of any way how to achieve this, however we believe this difference to be of the order of $|\sum_{\bsigma \in \Sigma_N}h(\bsigma)\pi(\bsigma,\infty) -  \int_{S_N} d\bsigma h(\bsigma) Q(\bsigma,\infty)| = \O{-\epsilon}$, which comes from the ratio of the second and fourth moment contributions. This is because after setting the time derivative to 0 and using the estimates of the derivatives of $h$, we obtain a perturbative form of the Fokker-Planck equation, i.e. $L_{\rm FP}Q = xQ$, where $L_{\rm FP}$ denotes the operator on the RHS of (\ref{FP eqn finally}) and $x = \O{-\epsilon}$. A precise error estimate will necessitate a careful stability analysis of the Fokker-Planck equation, which is beyond the scope of the present article.

\section{Real symmetric Bernoulli ensemble}\label{Real symmetric}
Our first example consists of the following matrix ensemble $\mathcal{B}_N$, in which the matrices $\bar{B} \in \mathcal{B}_N$ have elements
\begin{equation}\label{Real Bernoulli}
\bar{B}_{pq} =
\left\{\begin{array}{ll}
\pm \sqrt{1/N}  & p<q  \\
\bar{B}_{qp} & p>q \\
\pm \sqrt{2/N}  & q=p ,
\end{array}\right.
\end{equation}
chosen with equal probability. The addition of the diagonal elements in comparison to the balanced matrices (\ref{eq:balanced}) serves to simplify the proceeding calculations but is not crucial for the result. It does mean the number of free parameters, and therefore the dimension of the associated hypercube is given by $d_N = N(N+1)/2$. Moreover, as remarked in Section \ref{fokker-planck}, one can verify that $\Tr(\bar{B}^2) = 2d_N/N = \mathcal{O}(N)$ for all $\bar{B} \in \mathcal{B}_N$. The ensemble (\ref{Real Bernoulli}) falls into the class of Wigner matrices and hence one can show that in the limit of large $N$, the mean spectral density of the ensemble approaches that of the Wigner semicircle $\rho(\lambda) = \sqrt{4 - \lambda^2}/(2\pi)$ (see e.g. \cite{zeitouni} for a definition of Wigner matrices and proof of this fact).

\subsection{Evaluation of moments}\label{moments}

Let us suppose that we have two matrices $\bar{B}' \sim \bar{B}$ that are neighbours in our meta-graph. $\bar{B}'$ is thus obtained from $\bar{B}$ by switching a single element or, equivalently, by adding a rank 2 (or rank 1 if it is a diagonal element) matrix $\delta \bar{B}$ to $\bar{B}$, given by
\begin{equation}\label{Matrix perturbation}
\delta \bar{B}^{(pq)} = \bar{B}' - \bar{B} =
\left\{ \begin{array}{ll}
-2\bar{B}_{pq}(|p\rangle \langle q| + | q \rangle \langle p|) & p<q \\
-2\bar{B}_{pq}|p\rangle \langle p| & p=q. \end{array}\right.
\end{equation}
Here we use the ket notation $|p\rangle$ to denote the column vector with a 1 in the $p$-th component and 0  everywhere else, and the bra $\langle p|$ denotes the transpose of $|p\rangle$. Importantly this perturbation depends on the original matrix $\bar{B}$, and therefore so will $\Delta \bar{B}$ in (\ref{Perturbation - multi-step}), provided that our random walker does not move back and forth in the same direction of the hypercube. Or, in other words, all $\omega_r$ belonging to the path $\omega = (\omega_1,\ldots,\omega_{\Delta t})$ are different. However, this situation occurs with probability 1 in the large $N$ limit (see Subsection \ref{Subsec: Linear terms}), since the number of possible choices $d_N$, far outweighs the number of steps of the random walker $\Delta t$.

The perturbation (\ref{Perturbation - multi-step}) in turn induces a change in each of the eigenvalues $\Delta \lambda_{\nu}$. Unfortunately this change cannot be calculated exactly but it can be approximated using the well-known perturbation formula
\begin{equation}\label{Perturbation formula}
\Delta \lambda_{\nu} = \langle \nu |\Delta \bar{B} |\nu\rangle +\sum_{\mu:\mu \ne \nu} \frac{|\langle  \nu |\Delta \bar{B} |\mu\rangle|^2}{\lambda_{\nu}-\lambda_{\mu}} + {\rm higher \; order \; terms} \; ,
\end{equation}
with the third order term given in (\ref{Third order perturbation}). In complete analogy with Dyson's Brownian motion, our aim is to use this expansion in order to find expressions for the moments (\ref{moments}), which will come from the linear and quadratic terms in the matrix perturbation $\Delta \bar{B}$. The difference is that we are dealing with a discrete process and therefore we must be careful regarding higher terms in the expansion, whereas in Dyson's case the higher terms are justifiably neglected as the time differential can be made arbitrarily small.  Nevertheless, (\ref{Perturbation - multi-step}) gives us an indication why this might be viable. Looking closely we see that, even though the elements of the matrix $\Delta \bar{B}$ will only take one of two values, each of the individual random perturbations $\delta \bar{B}$ are (almost) independent and identically distributed. Therefore, by reducing to the space of spectra, terms such as $\langle \nu |\Delta \bar{B} |\nu\rangle$ will become a sum of iid random variables and hence, due to the CLT will approach a Gaussian distribution with appropriate mean and a variance proportional to $\Delta t$, which is the essence of Brownian motion (see e.g. \cite{Wang,Risken-1989}).

This observation also imposes a restriction on the range of $\Delta t$ for which the perturbation formula provides a good approximation to $\Delta \lambda_{\nu}$. Since $ \langle \nu |\Delta \bar{B} |\nu\rangle$ is the main term in this approximation, we require this to be much less than the typical spacing between eigenvalues, given by $\bar{\Delta} = 4/N$ . Therefore, since the variance grows linearly in $\Delta t$ the standard deviation is
\[
\frac{\sqrt{\Var{\langle \nu |\Delta \bar{B} |\nu\rangle}}}{\bar{\Delta}} \approx \frac{N}{4} \sqrt{\Delta t} \sqrt{\E(\langle \nu |\Delta \bar{B} |\nu\rangle^2)} \approx \sqrt{\frac{\Delta t}{N}},
\]
where we have used that $\E(\langle \nu |\Delta \bar{B} |\nu\rangle^2) = 8/(Nd_N)$ (see Subsection \ref{Subsec: Quadratic terms}) and $d_N \approx N^2/2$. Thus we require $\Delta t \ll N$, which is comfortably satisfied if we take $\Delta t = \O{c}$ with $c<1$. We shall show below this same restriction also emerges from different reasons.

In the following we shall calculate the expectation of terms both linear and quadratic in the matrix perturbation $\Delta \bar{B}$, for example $\langle \nu |\Delta \bar{B} |\nu\rangle$ and $\langle  \nu |\Delta \bar{B} |\mu\rangle^2$. These will be calculated explicitly and will provide support for the line of thought discussed in the introducing remarks to section \ref{Spectral statistics} - concerning the relation to the CLT. Collating these will allow us to obtain expressions for the first two moments $M_{\nu}$ and $M_{\nu\mu}$ (see (\ref{moments definition})) which in turn allow us to obtain a Fokker-Plank equation and therefore the stationary JPDF. This is presented in Subsection \ref{Subsec: Stat sol}. The estimation of higher orders, specifically cubic and quartic terms, are left until \ref{Higher moments}.

\subsubsection{Linear terms}\label{Subsec: Linear terms}

We begin by calculating the expectation of the first term $\langle \nu |\Delta \bar{B} |\nu\rangle$ in (\ref{Perturbation formula}) due to a random walk of $\Delta t$ time steps. Given the linearity of this expression we can directly calculate the expectation in the matrix perturbation, $\E(\Delta \bar{B})$, which, using (\ref{Transition probability}), is given by
\begin{equation}
\fl \E(\Delta \bar{B}) = \sum_{\bar{B}'} \rho_{\Delta t}(\bar{B}\to \bar{B}')\Delta \bar{B}(\bar{B}')
=  \sum_{X=0}^{\Delta t}  \frac{P_{\Delta t}(X)}{\mathcal{N}_X}\sum_{\bar{B}' : \bar{D}(\bar{B},\bar{B}') = X} \Delta \bar{B}(\bar{B}'). \end{equation}
Here $P_{\Delta t}(X)$ is the transition probability to go from any matrix $\bar{B}$, to another $\bar{B}'$ a distance $X$ away, as described in Section \ref{entropic forces}. The probability to reach all such matrices is the same and thus we simply divide by their number $\mathcal{N}_X$. Alternatively, this can be viewed as summing over the set $\Omega_X$ of all ordered paths of length $X$. These are paths $\omega = (\omega_1,\ldots,\omega_X)$ in which all $\omega_r$ are different and arranged in, say, decreasing order and each such $\omega \in \Omega_X$ corresponds to a unique $\bar{B}'$, a distance $X$ away from $\bar{B}$. Hence $|\mathcal{N}_X| = |\Omega_X|$ and
\begin{equation}\label{Perturbation expectation}
\E(\Delta \bar{B}) = \sum_{X=0}^{\Delta t}  \frac{P_{\Delta t}(X)}{|\Omega_X|} \sum_{\omega \in \Omega_X} \Delta \bar{B}^\omega = \sum_{X=0}^{\Delta t} \frac{P_{\Delta t}(X)}{|\Omega_X|} \sum_{\omega \in \Omega_X} \sum_{r = 1}^X \delta \bar{B}^{\omega_r},
\end{equation}
where we have inserted the summation over single step perturbations (\ref{Perturbation - multi-step}) and, in comparison with the stationary distribution (\ref{Stat distribution Gaussian}),
\begin{equation}\label{Ordered paths}
|\Omega_X| = \choose{d_N}{X}.
\end{equation}
The double sum in (\ref{Perturbation expectation}) can be written as a single sum over the nearest neighbours of $\bar{B}$, i.e. $\sum_{\omega \in \Omega_X} \sum_{r = 1}^X \delta \bar{B}^{\omega_r} = \sum_{p \leq q} \Phi^{(pq)}_X \delta \bar{B}^{pq}$, where $\Phi^{(pq)}_X$ denotes the number of paths $\omega \in \Omega_X$ that contain the element $pq$. However, this must be the same for all $(pq)$, so without loss of generality we can write $\Phi^{(pq)}_X = \Phi^{(1)}_X$, which is given by conditioning $(pq)$ to belong to $\omega$ and choosing the remaining $X-1$ elements from the $d_N-1$ possible options, i.e.
\begin{equation}\label{fixed ordered paths}
\Phi^{(1)}_X = \choose{d_N - 1}{X - 1} = \frac{X}{d_N}\choose{d_N}{X} = \frac{X}{d_N}|\Omega_X|.
\end{equation}
Therefore removing $\Phi^{(1)}_X$ outside the sum we obtain
\begin{equation}
\fl
\E(\Delta \bar{B}) = \sum_{X=0}^{\Delta t} \frac{P_{\Delta t}(X)}{|\Omega_X|} \Phi_X^{(1)} \sum_{p \leq q} \delta \bar{B}^{pq} = \left[\frac{1}{d_N}\sum_{X=0}^{\Delta t} X P_{\Delta t}(X)\right]\left[ \sum_{p \leq q} \delta \bar{B}^{pq}\right].
\end{equation}

We are left with two seperate contributions to determine; one from the expected distance of the random walker away from the initial matrix $\bar{B}$ after $\Delta t$ time steps, the other from the aforementioned nearest neighbour sum. Let us start with the former. We know this cannot exceed $\Delta t$, therefore we can write
\[
\Delta t P_{\Delta t}(\Delta t) < \sum_{X=0}^{\Delta t} X P_{\Delta t}(X) < \Delta t
\]
for all $\Delta t > 0$. It thus suffices to estimate $P_{\Delta t}(\Delta t)$ for $\Delta t \ll d_N$. From Section \ref{random walks} we know that at each time step the random walker has $(d_N+1)$ options to either stay in the same place, or move to an immediate neighbour. Thus after $\Delta t$ steps there is a total of $(d_N+1)^{\Delta t}$ possible paths the random walker can take, with $\Delta t! |\Omega_{\Delta t}|$ of these options leading the random walker to a matrix $\bar{B}'$ that is a distance $\Delta t$ away from $\bar{B}$.
\begin{equation}\label{RW estimate}
\fl
P_{\Delta t}(\Delta t) = \frac{\Delta t !}{(d_N+1)^{\Delta t}}\choose{d_N}{\Delta t} = \prod_{j=0}^{\Delta t-1}\frac{d_N - j}{(d_N+1)^{\Delta t}} = 1 - \frac{1}{d_N}\sum_{j=0}^{\Delta t-1} j + \ldots.
\end{equation}
Hence, for $\Delta t \ll d_N$, we have $P_{\Delta t}(\Delta t) = 1 + \mathcal{O}(\Delta t^2/d_N) = 1 + \O{2c-2}$ and so
\[
\E(\Delta \bar{B}) = (1 + \O{2c-2})\frac{\Delta t}{d_N}\sum_{p \leq q} \delta \bar{B}^{pq}.
\]

This leaves us with the second contribution from the nearest neighbour sum. Using the expression for the single step perturbation (\ref{Matrix perturbation}) we have
\begin{equation}\label{NN sum}
\sum_{p\leq q} \delta \bar{B}^{(pq)} = - 2\sum_{p < q} \bar{B}_{pq} (|p\rangle \langle q | + |q\rangle \langle p|) - 2\sum_p \bar{B}_{pp} |p\rangle \langle p | = -2\bar{B}.
\end{equation}
Therefore
\begin{equation}\label{expectation}
\frac{\E(\langle \nu | \Delta \bar{B} | \nu \rangle)}{\Delta \eta} = -2\langle \nu | \bar{B} | \nu \rangle(1 + \O{2c-2}) = -2\lambda_\nu + \O{2c-2},
\end{equation}
where $\Delta \eta = \Delta t/d_N$ and $\lambda_{\nu} = \mathcal{O}(1)$. Up to a correction, this result is exactly what one would expect for the Gaussian ensembles, showing there is an overall `force' on the spectrum that acts to bring the eigenvalues closer to 0. However in contrast to Dyson's case \cite{Dyson-1962} it arises not because of an {\it imposed}  Gaussian distribution of the matrix variations,  but because of the entropic forces experienced by the random walker as it moves within the hypercube, just as in Section \ref{entropic forces}. The restriction of $c$ to the interval $0<c<1$ is necessary for the error term in (\ref {expectation}) to tend to $0$ when $N\rightarrow \infty$.

\subsubsection{Quadratic terms}\label{Subsec: Quadratic terms}

We now turn our attention to those instances in which we have contributions  quadratic in $\Delta \bar{B}$. There are two such scenarios. The first occurs in the second term of the perturbation formula (\ref{Perturbation formula}), the second in the leading contribution to $M_{\mu\nu}$. The former concerns quantities $\E(|\langle \nu |\Delta \bar{B}|\mu \rangle |^2)$ with $\nu \neq \mu$, whereas the latter those of the type $\E(\langle \nu |\Delta \bar{B} | \nu\rangle\langle \mu |\Delta \bar{B} | \mu\rangle)$. We start by evaluating the first case in the same manner as for the linear terms. The expectation is therefore given, analogously to (\ref{Perturbation expectation}), as
\[
\E(|\langle \nu | \Delta \bar{B} | \mu \rangle |^2)   = \sum_{X=0}^{\Delta t} \frac{P_{\Delta t}(X)}{|\Omega_X|} \sum_{\omega \in \Omega_X} \sum_{r,s = 1}^X \langle \nu | \delta \bar{B}^{\omega_r}| \mu \rangle \langle \mu | \delta \bar{B}^{\omega_s}| \nu \rangle.
\]
where, we have used the expression (\ref{Perturbation - multi-step}) for the multi-step perturbation of $\Delta \bar{B}$. Here we must now split the double sum in terms of diagonal and off-diagonal parts, i.e. $\sum_{r,s} = \sum_r + \sum_{r \neq s}$. The reason being that the diagonal term, as before, consists of counting those ordered paths $\omega$ that contain a particular element $(pq)$, whereas in the off-diagonal term we must count the number that contain two distinctive elements. In fact the expression (\ref{fixed ordered paths}) generalises quite straightforwardly to determine the number of ordered paths of length $X$ with $l$ distinct elements, to
\begin{equation}\label{fixed ordered paths 2}
\Phi^{(l)}_X = \choose{d_N - l}{X - l} = |\Omega_X| \prod_{i=0}^{l-1} \frac{X-i}{d_N-i}.
\end{equation}
Using this we find
\begin{equation}\label{Quadratic expansion}
\fl
\E(|\langle \nu | \Delta \bar{B} | \mu \rangle |^2)   = \sum_{X=0}^{\Delta t} \frac{P_{\Delta t}(X)}{|\Omega_X|}
\left[\Phi^{(1)}_X\sum_e |\langle \nu | \delta \bar{B}^{e}| \mu \rangle|^2 + \Phi^{(2)}_X\sum_{e \neq e'}\langle \nu | \delta \bar{B}^{e}| \mu \rangle \langle \mu | \delta \bar{B}^{e'}| \nu \rangle\right],
\end{equation}
where $e$ denotes one of the $d_N$ independent elements $pq$ of the matrix. The diagonal term can be evaluated in the same way as before, to get
\begin{equation}\label{Diagonal quadratic term}
\fl
\frac{1}{d_N}\sum_{X=0}^{\Delta t} XP_{\Delta t}(X) \sum_{p \leq q} |\langle \nu | \delta \bar{B}^{pq}| \mu \rangle|^2 = \frac{\Delta t}{d_N}  \sum_{p \leq q} |\langle \nu | \delta \bar{B}^{pq}| \mu \rangle|^2(1 + \O{2c - 2}).
\end{equation}
Summing over the nearest neighbours then gives
\begin{eqnarray}
\fl
 \sum_{p \leq q} |\langle \nu | \delta \bar{B}^{pq}| \mu \rangle|^2 & = & \sum_{p<q}|\bar{B}_{pq}\langle \nu |p\rangle\langle q |\mu \rangle + \bar{B}_{qp}\langle \nu |q\rangle\langle p| \mu \rangle|^2  + \sum_p \bar{B}_{pp}^2 |\langle \nu |p\rangle \langle p| \mu \rangle|^2  \nonumber \\
& = & \frac{4}{N}\left[\sum_{p<q}(\nu_p^2 \mu_q^2  + 2 \nu_p \mu_q \nu_q \mu_p + \nu_q^2 \mu_p^2)  + \sum_p 2 \nu_p^2\mu_p^2\right] \nonumber \\
& = & \frac{4}{N}\sum_{p,q}(\nu_p^2 \mu_q^2  + \nu_p \mu_q \nu_q \mu_p) = \frac{4}{N},
\end{eqnarray}
where we have used that $\nu \neq \mu$. The off-diagonal term gives the following correction
\[
\fl \frac{\Delta t(\Delta t - 1)}{d_N(d_N - 1)}\left[ \left|\sum_{p\leq q} \langle \nu | \delta \bar{B}^{(pq)} | \mu \rangle \right|^2 - \sum_{p \leq q} |\langle \nu | \delta \bar{B}^{(pq)} | \mu \rangle|^2 \right] =  \mathcal{O}(\Delta \eta^2) \frac{4}{N},
\]
where again, we have used that $\nu \neq \mu$ and the prefactor comes from $\sum_{X=0}^{\Delta t} P_{\Delta t}(X)\Phi^{(2)}_{X}/|\Omega_X| \approx P_{\Delta t}(\Delta t)\Phi^{(2)}_{\Delta t}/|\Omega_{\Delta t}|$. However this correction is less than the one arising in the diagonal term (\ref{Diagonal quadratic term}), thus
\begin{equation}\label{variance}
\frac{\E(|\langle \nu | \Delta \bar{B} | \mu \rangle |^2)}{\Delta \eta} = \frac{4}{N}(1 + \O{2c-2}) = \frac{4}{N} + \O{2c-3}.
\end{equation}

\vspace{20pt}

The second scenario involves terms of the form $\E(\langle \nu |\Delta \bar{B} | \nu\rangle\langle \mu |\Delta \bar{B} | \mu\rangle)$. Therefore, starting once more from (\ref{Perturbation expectation})
\[
\fl
\E(\langle \nu |\Delta \bar{B} | \nu\rangle\langle \mu |\Delta \bar{B} | \mu\rangle)  = \sum_{X=0}^{\Delta t} \frac{P_{\Delta t}(X)}{|\Omega_X|} \sum_{\omega \in \Omega_X} \sum_{r,s = 1}^X \langle \nu | \delta \bar{B}^{\omega_r}| \nu \rangle \langle \mu | \delta \bar{B}^{\omega_s}| \mu \rangle
\]
and proceeding as in the former case we get a diagonal term
\[
\Delta \eta \sum_{p \leq q}  \langle \nu | \delta \bar{B}^{pq}| \nu \rangle \langle \mu | \delta \bar{B}^{pq}| \mu \rangle(1 + \O{2c -2}),
\]
where it can be shown that
\[
 \sum_{p \leq q}  \langle \nu | \delta \bar{B}^{pq}| \nu \rangle \langle \mu | \delta \bar{B}^{pq}| \mu \rangle = \frac{8}{N}\delta_{\mu\nu}.
\]
Similarly, going through the steps above the off-diagonal part is given by
\[
\fl
\mathcal{O}(\Delta \eta^2) \left[ \sum_{p\leq q} \langle \nu | \delta \bar{B}^{(pq)} | \nu \rangle\langle \mu | \delta \bar{B}^{(pq)} | \mu \rangle - \left(\sum_{p \leq q} \langle \nu | \delta \bar{B}^{(pq)} | \nu \rangle\right) \left(\sum_{p \leq q} \langle \mu | \delta \bar{B}^{(pq)} | \mu \rangle\right) \right],
\]
which equals
\[
\mathcal{O}(\Delta \eta^2)\left[ \frac{8}{N}\delta_{\nu\mu} - 4\lambda_{\nu}\lambda_{\mu}\right] = \Delta \eta \O{c-3},
\]
using $\lambda_\nu = \mathcal{O}(1)$. Therefore
\begin{equation}\label{Quadratic result}
\frac{\E(\langle \nu |\Delta \bar{B} | \nu\rangle\langle \mu |\Delta \bar{B} | \mu\rangle)}{\Delta \eta} =
\left\{ \begin{array}{ll}
\frac{8}{N} + \O{2c-3} & \nu = \mu \\
\O{c - 3} & \nu \neq \mu \; . \end{array}\right.
\end{equation}

The results for the linear and quadratic terms from Subsections \ref{Subsec: Linear terms} and \ref{Subsec: Quadratic terms}, together with calculations in \ref{App: Cubic terms} and \ref{App: Quartic terms}, provide expressions for the first two moments, together with error estimates (which rely on results concerning the delocalisation of eigenvectors (see e.g. \cite{Tao-2011b,Erdos-2011,Erdos-2013} and references therein). In particular, for the first order moments we find
\begin{equation}\label{Moment result 1}
M_{\nu} = -2\lambda_\nu + \frac{4}{N}\sum_{\mu:\mu \neq \nu}\frac{1}{\lambda_\nu - \lambda_{\mu}} + \O{c+q-1}.
\end{equation}
Here the correction comes from the third order term in the perturbation expansion (\ref{Third order perturbation}). The numerator of which is estimated in \ref{App: Cubic terms}, whilst the denominator we estimate using the spectral rigidity properties of Wigner matrices provided in \cite{Tao-2011b}. This says that the spacing between eigenvalues is $\O{-1-q}$ with high probability, where $q$ is some small positive constant. Similarly, the main correction to the moments $M_{\nu\nu}$ arises from expanding out the perturbation formula (\ref{Perturbation formula}) and evaluating the cubic terms in the following
\[
\sum_{\mu: \mu \neq \nu} \frac{\E(\langle \nu |\Delta \bar{B}| \nu\rangle\langle \nu |\Delta \bar{B}| \mu\rangle \langle \mu |\Delta \bar{B}| \nu\rangle  )}{\lambda_{\nu} - \lambda_{\mu}}.
\]
Again we may use results in \ref{App: Cubic terms}, which show the numerator is of the order of $\O{c-3}$, bringing us to
\begin{equation}\label{Moment result 2}
M_{\nu\nu} = \frac{8}{N} + \O{c+q-2}.
\end{equation}

\subsection{Stationary solution}\label{Subsec: Stat sol}
If we take the leading order expressions for the moments (\ref{Moment result 1}) and (\ref{Moment result 2}) and insert these into (\ref{FP eqn finally}),  we obtain the same Fokker-Planck equation as Dyson \cite{Dyson-1962}, which describes the stochastic motion of the eigenvalues in the GOE ensemble. However in our case we must condition the eigenvalues to obey a fixed trace, since $\Tr(\bar{B}^2) = d_N/N$. Hence, setting the time-derivative to zero and solving for the stationary distribution $Q(\bsigma) \equiv Q(\bsigma,\infty)$ we recover the JPDF for the fixed trace Gaussian ensemble \cite{Mehtabook}
\begin{equation}
Q(\bsigma) = C_N \prod_{\mu < \nu}| \lambda_{\nu}-\lambda_{\mu}| \ \delta\left(\sum_{\nu=1}^N\lambda_{\nu}^2 - 2d_N/N\right),
\end{equation}
where $C_N$ is an appropriate normalisation constant. The properties of this ensemble are studied in various articles, where it is proven that the mean spectral density approaches the semi-circle law, and that the deviations for finite $N$ follow the Tracey-Widom theory \cite{Zhoua-2010}.

\section{Imaginary anti-symmetric Bernoulli ensemble}\label{Imaginary anti-symmetric}

In our second example we investigate Hermitian Bernoulli matrices $H$ with purely imaginary elements. If one takes the adjacency matrix of the tournaments graphs outlined in point (\ref{Tournament graphs}) in the introduction then we have a direct relation via the transformation $H = i\bar{B} = i(2A - J + I)/sqrt{N}$.
\begin{equation}\label{Imaginary Bernoulli}
H_{pq} =
\left\{\begin{array}{ll}
\pm i/\sqrt{N}  & p<q \\
-H_{qp} & p>q \\
0 & q=p \; .
\end{array}\right.
\end{equation}
We would now like to discern the stochastic motion in the eigenvalues of the ensemble of $H$ and from this extract the JPDF for the eigenvalues. In this instance all the matrices satisfy the anti-commutation relation $H^* = -H$, which implies the eigenvalues come in pairs $\{ \pm \lambda_{\nu}\}$ with associated eigenvectors $|\nu\rangle$ and $|\nu^*\rangle$ that are complex conjugates of each other. This means if $N$ is odd there is a single eigenvalue denoted by $\lambda_0$ equal to 0. Note also the slight difference in this scenario that, due to the lack of diagonal elements, $d_N = N(N-1)/2$.

We would also like to mention that recently, it has been proven in \cite{Sosoe}, using a different approach based upon the work in \cite{Erdos-2010b,Erdos-2010c,Erdos-2012c,Erdos-2012d}, that all order correlation functions for the eigenvalues of $H$ converge weakly to that of the associated Gaussian ensemble JPDF of (\ref{Anti-symmetric JPDF}).

\subsection{Evaluation of Moments}
The evaluation of the moments follows in exactly the same manner as in Section \ref{moments} and therefore many details will be omitted. Using the estimate of the random walker a distance $\Delta t$ away from the origin (\ref{RW estimate}) we once again find for the linear terms in $\Delta H$
\[
\E(\Delta H) = (1 + \mathcal{O}(\Delta \eta))\Delta \eta \sum_{p  < q} \delta H^{pq}.
\]
Here the change in the matrix $H$ induced by switching the $pq$-th element is $\delta H^{pq} =-2H_{pq} [|p\rangle\langle q| - |q\rangle\langle p|]$ is slightly different from the example in Section \ref{Real Bernoulli}. However after summing over all elements we recover an analogous expression to (\ref{NN sum})
\[
\sum_{p < q} \delta H^{pq} = -2 \sum_{p < q} H_{pq} [|p\rangle\langle q| - |q\rangle\langle p|] = -2H.
\]
Therefore,
\[
\frac{\E(\langle \nu | \Delta H | \nu \rangle)}{\Delta \eta} = -2\lambda_\nu + \mathcal{O}(N^{2c - 2}).
\]

For the quadratic terms we recover an analogous expansion (\ref{Quadratic expansion}), with both diagonal (\ref{Diagonal quadratic term}) and off-diagonal terms. The former of which gives a leading contribution of the form
\[
\E(|\langle \nu | \Delta H | \mu \rangle |^2) = \Delta \eta(1 + \O{2c-2})\sum_{p < q}|\langle \nu | \delta H^{pq} | \mu \rangle |^2,
\]
with
\[
\fl
\sum_{p < q}|\langle \nu | \delta H^{pq} | \mu \rangle |^2  = \frac{4}{N} \sum_{p,q}(|\nu_p|^2|\mu_q|^2 - \nu_p \nu_q^* \mu_p \mu_q^*)
=  \left\{\begin{array}{ll}
4/N & \nu \neq \mu^*\\
0 & \nu = \mu^*. \end{array}\right.
\]
Here we have used that $|H_{pq}|^2 = 1/N$ and since $|\nu \rangle$ and $|\nu^* \rangle$ are orthogonal eigenvectors $\sum_p \nu_p^{2} = \sum_p \nu_p^{*2} = 0$. Therefore, using the perturbation formula (\ref{Perturbation formula}), we have, very similar to the previous section,
\begin{equation}\label{AS moment 1}
M_{\nu} = -2\lambda_\nu + \frac{4}{N}\sum_{\mu:\mu \neq \nu,\nu^*} \frac{1}{\lambda_{\nu} - \lambda_{\mu}} + \O{c+q -1}
\end{equation}
and
\begin{equation}\label{AS moment 2}
M_{\nu\nu} = \frac{4}{N} + \O{c+q -2}.
\end{equation}
The errors once again appear from higher terms in the perturbation formula, which require knowledge about eigenvector delocalisation and eigenvalue rigidity.

If again we take the leading order expressions in (\ref{AS moment 1}) and (\ref{AS moment 2}) and insert these into the Fokker-Planck equation (\ref{FP eqn main}) we can obtain an expression for the JPDF $Q(\bsigma) \equiv Q(\bsigma,\infty)$. In contrast to the previous section, the eigenvalues come in pairs $\{\pm \lambda_{\nu}\}$, which we group together. This means if $N$ is odd then we have the labels $\nu = 1,\ldots,(N-1)/2$ so the expression for the first moment can be written as
\[
M_{\nu} =  -2\lambda_\nu + \frac{4}{N}\sum_{\mu:\mu\neq \nu} \left[\frac{1}{\lambda_\nu - \lambda_\mu}  + \frac{1}{\lambda_\nu + \lambda_\mu}\right] + \frac{4}{N\lambda_\nu},
\]
where the last term comes from the $\lambda_0=0$ eigenvalue. The stationary solution to the Fokker-Planck equation (\ref{fokker-planck}) is given by writing $Q(\bsigma) = e^{-V(\bsigma)}$. Doing so we reach the equation for the force
\[
\frac{\partial V(\bsigma)}{\partial \lambda_\nu} = -2\frac{M_{\nu}}{M_{\nu\nu}} =
\left[N\lambda_\nu - \frac{2}{\lambda_\nu} - \sum_{\mu:\mu\neq \nu} \frac{4\lambda_n }{\lambda^2_\nu - \lambda^2_\mu}\right].
\]
Integrating over $\lambda_\nu$ and summing over $\nu$ we obtain for the potential
\[
V(\bsigma) = \sum_\nu \left( \frac{N\lambda_\nu^2}{2} - 2\ln(\lambda_\nu) \right) - 2\sum_{\nu < \mu} \ln |\lambda_\nu^2 - \lambda_\mu^2|,
\]
which equates to the following JPDF for the eigenvalues.
\begin{equation}\label{Anti-symmetric JPDF}
Q(\bsigma) = C_N\prod_{\nu}\lambda_\nu^2\prod_{n<m} |\lambda^2_m - \lambda^2_n |^2\delta\left(\sum_{\nu}\lambda_\nu^2 - d_N/N\right)
\end{equation}
This corresponds to the fixed trace symmetric GUE ensemble (see Section 3.4 of \cite{Mehtabook}).

\section{Real Wishart Bernoulli ensemble}\label{Real Wishart}
As a final example, we turn our attention to Bernoulli matrices $\bar{B}$ of size $N \times M$ (we shall assume $N \geq M$ without loss of generality). Once again, every element is chosen independently from the set $\{ \pm 1\}/\sqrt{N}$ with equal probability and we denote this ensemble $\mathcal{B}_N$ as before. Such matrices are related to ensembles of random undirected graphs, as outlined in point (\ref{Directed graphs}) of the introduction. In the case $N \neq M$ the matrix $\bar{B}$ does not have any eigenvalues, however there exists a factorisation of the form
\[
\bar{B} = U\Lambda V^T
\]
where $\Lambda$ is an $N \times M$ matrix of \emph{singular values} $s_\nu$ on the diagonal and the columns of the $N \times N$ matrix $U$ and $M \times M$ matrix $V$ give the left and right \emph{singular vectors} respectively. The eigenvalues $\lambda_{\nu}$ of the matrix
\begin{equation}\label{SV matrix}
W = \bar{B}^T\bar{B}.
\end{equation}
are the square of the singular values, i.e. $\lambda_{\nu} = s_{\nu}^2$ with (real) eigenvectors $S | \nu \rangle = \lambda_{\nu}| \nu \rangle$ provided by the columns of $V$. Due to the nature of $\bar{B}$, $W$ commands a fixed trace $\Tr(W) = M$ for all $\bar{B} \in \mathcal{B}_N$.

As in the previous two examples, our Bernoulli ensemble has a Gaussian equivalent, known as the Wishart-Laguerre ensemble - corresponding to matrices $W$ when the elements of $\bar{B}$ are normally distributed. In this case the JPDF for the eigenvalues can be determined exactly for any $N$, $M$, either by orthogonal polynomial techniques \cite{Bronk-1965,Forrester-1993}, or via an appropriately adapted version of Dyson's Brownian motion approach \cite{Bru-1989,Bru-1991,Akuzawa-1997}. Analysis has also been performed for the eigenvalues of fixed-trace Wishart-Laguerre ensembles (see e.g. \cite{Akemann-2011} and references therein).

We will be interested in the limit $N \to \infty$, with the ratio $M/N \to c < 1$. In this case it is known that the mean spectral density for our Bernoulli ensemble tends to the following distribution
\begin{equation}\label{Marchenko-Pastur}
\rho(\lambda) = \frac{1}{2\pi c}\frac{\sqrt{(a_+ -\lambda)(\lambda - a_-)}}{\lambda},
\end{equation}
where $a_{\pm} = (1 \pm \sqrt{c})^2$ and is supported in the interval $[a_-,a_+]$. This is the Marchenko-Pastur law \cite{Marchenko-1967} and implies our mean level spacing is of order $\O{-1}$. We are interested in the fine scale behaviour of the eigenvalues of matrices (\ref{SV matrix}). Recently this has been addressed in \cite{Tao-2012}, where the authors develop techniques used to treat Wigner matrices \cite{Tao-2011a}, in order to analyse matrices $\bar{B}$ with elements of quite general probability distribution.

\vspace{10pt}
Once again, we want to set up a random walk through $\mathcal{B}_N$, analogous to that in Section \ref{Real symmetric}. The matrices $\bar{B}$ also represent the corners of a $d_N$-dimensional hypercube but this time with $d_N = NM$, the number of independent parameters in $\bar{B}$. A Hamming distance of 1 on the hypercube therefore corresponds to swapping just one element $\bar{B}_{pq}$, which means the single step perturbation $\delta \bar{B} = \bar{B}' - \bar{B}$ is the rank 1 matrix
\begin{equation}\label{Wishart perturbation}
\delta \bar{B}^{(pq)} = -2\bar{B}_{pq}|p \rangle \langle q|.
\end{equation}
Similarly, a $\Delta t$-step perturbation $\Delta \bar{B}$, given by (\ref{Perturbation - multi-step}), leads to the following perturbation in $W$
\begin{equation}\label{Wishart expansion}
\fl
W + \Delta W = (\bar{B} +\Delta \bar{B})^T(\bar{B} +\Delta \bar{B}) = W + [\bar{B}^T\Delta \bar{B} + \Delta \bar{B}^T \bar{B} + \Delta \bar{B}^T \Delta \bar{B}].
\end{equation}
The task is to evaluate the moments of the spectral evolution due to a change in the matrix of $\Delta W$. The details of which shall be omitted once again, since they follow very closely what has been outlined already in Section \ref{Real symmetric}. Like in the previous two cases errors will inevitably be incurred because of the finite time effects of our random walk. These can be estimated thanks to delocalisation estimates of the eigenvectors which are applicable in covariance matrices \cite{Tao-2012,Rudelson-2014} and results concerning spectral rigidity \cite{Tao-2012} for Wigner matrices.

\subsection{Evaluation of moments}
We begin with the expectation $\E(\Delta W)$ of the change in the matrix due to a random walk of $\Delta t$-steps. Using the expansion in terms of $\Delta \bar{B}$ given by (\ref{Wishart expansion}) and the analogous expression (\ref{Perturbation expectation}) we find
\begin{equation}\label{Wishart expectation}
\fl
\E(\Delta W)  = \sum_{X=0}^{\Delta t} \frac{P_{\Delta t}(X)}{|\Omega_X|}\sum_{\omega \in \Omega_X} \left[\sum_{r \in \omega}[ \bar{B}^T \delta \bar{B}^{\omega_r} + (\delta \bar{B}^{\omega_r})^T\bar{B}] + \sum_{r,s \in \omega}(\delta \bar{B}^{\omega_r})^T\delta \bar{B}^{\omega_s}\right].
\end{equation}
In contrast to the previous two cases this contains quadratic terms in the matrix elements of $\delta \bar{B}$, however this only presents an additional source of error that can be calculated. The leading term comes from a combination of the linear terms and the diagonal terms in the quadratic part, this is
\[
\E(\Delta W) = \Delta \eta\sum_{p,q=1}^{N,M} \left[\bar{B}^T\delta \bar{B}^{pq} + (\delta \bar{B}^{pq})^T\bar{B} +  (\delta \bar{B}^{pq})^T(\delta \bar{B}^{pq}) \right].
\]
Inserting the rank 1 matrix (\ref{Wishart perturbation}), summing over $p$ and $q$ and applying this to the first term in the perturbation formula (\ref{Perturbation formula}) we get
\begin{equation}\label{Wishart first order}
\frac{\E(\langle \nu | \Delta W|\nu\rangle)}{\Delta \eta} = \langle \nu | 4(I_M - W) |\nu \rangle + \mathcal{O}(\Delta \eta) = 4(1 - \lambda_{\nu}) + \O{2c-2}.
\end{equation}
The leading correction in fact comes from the estimation of $P_{\Delta t}(\Delta t)$, as performed in (\ref{RW estimate}). The off-diagonal terms in the quadratic part of (\ref{Wishart expectation}) provide a correction of order $\mathcal{O}(\Delta \eta) = \O{c-2}$ which is slightly smaller.

\vspace{10pt}
If we proceed in a similar manner then we also obtain the second order contribution to the perturbation formula
\[
\frac{\E(|\langle \nu |\Delta W| \mu \rangle|^2)}{\Delta \eta} = (1 + \mathcal{O}(N^{2c-2}))\sum_{p,q} \langle \nu |\delta W^{pq}| \mu \rangle^2.
\]
Removing the absolute value is allowed since the inner product in real. In the previous two sections the analogous quantity obtained by summing over all matrix elements could be calculated exactly, whereas here we will will only get our desired expression up to some error. This is because $\delta W$ itself is comprised of quadratic terms in $\delta \bar{B}$, leading to cubic and quartic terms when $\delta W$ is squared, unlike the previous cases. Writing this out explicitly we find
\begin{eqnarray}
\fl
\sum_{p,q}\langle \nu |\delta W^{pq}| \mu \rangle^2 = & \sum_{p,q} & \left[\langle \nu |\bar{B}^T \delta \bar{B}^{pq} | \mu \rangle^2 +  \langle \nu |(\delta \bar{B}^{pq})^T \bar{B} | \mu \rangle^2 \right.\nonumber \\
& + & 2 \langle \nu |\bar{B}^T \delta \bar{B}^{pq} | \mu \rangle \langle \nu |(\delta \bar{B}^{pq})^T \bar{B} | \mu \rangle
\nonumber \\
& + & \langle \nu |(\delta \bar{B}^{pq})^T \delta \bar{B}^{pq} | \mu \rangle^2  + 2 \langle \nu |(\delta \bar{B}^{pq})^T  \bar{B} | \mu \rangle \langle \nu |(\delta \bar{B}^{pq})^T \delta \bar{B}^{pq} | \mu \rangle
\nonumber \\
 & + & \left.2 \langle \nu |\bar{B}^T\delta \bar{B}^{pq} | \mu \rangle \langle \mu |(\delta \bar{B}^{pq})^T \delta \bar{B}^{pq} | \mu \rangle \right].
\end{eqnarray}
Inserting the expression (\ref{Wishart perturbation}) and summing over $p$ and $q$ we find
\begin{equation}\label{Wishart second order}
\sum_{p,q}\langle \nu |\delta W^{pq}| \mu \rangle^2 = \frac{4}{N}[\lambda_{\nu} + \lambda_{\mu} + 2\delta_{\nu\mu}\lambda_{\nu}] + \O{-2},
\end{equation}
where the leading order expression comes from the first three terms, quadratic in $\delta \bar{B}$, and the remaining correction comes from the final three terms
\[
\left(\frac{4}{N}\right)^2(M - \lambda_{\nu} - \lambda_{\mu})\sum_{q}\nu_q^2\mu_q^2 = \O{-2}.
\]
Here we have used that $\sum_q \nu_q^2\mu_q^2 \approx M/M^2$, if they are delocalised, which we know happens with a high probability \cite{Tao-2012,Rudelson-2014}. Collating (\ref{Wishart first order}) and (\ref{Wishart second order}), together with the perturbation formula (\ref{Perturbation formula}) we obtain the leading order expressions for the first two moments
\begin{equation}\label{Wishart moments}
\fl
M_{\nu} = 4\left(1 - \lambda_{\nu}\right) + \frac{4}{N}\sum_{\mu: \mu \neq \nu} \frac{\lambda_{\nu} + \lambda_{\mu}}{\lambda_{\nu} - \lambda_{\mu}} \hspace{30pt} \mbox{and} \hspace{30pt} M_{\nu\nu} =  \frac{16}{N}\lambda_{\nu}.
\end{equation}
Again, the main corrections to these will come from higher order terms in the perturbation formula, which will rely on estimates regarding the rigidity of eigenvalues, recently obtained via local semicircle arguments \cite{Tao-2012}. One may also verify the off-diagonal, second order moments $M_{\nu\mu}$ are of a sufficiently lower order than the diagonal ones $M_{\nu\nu}$ and can be neglected in the large $N$ limit. Thus if the moments (\ref{Wishart moments}) are inserted into the Fokker-Planck equation the one may verify  the resulting stationary solution is given by
\[
Q(\bsigma) = \prod_\nu \lambda_{\nu}^{\frac{1}{2}(N - M + 1) - 1} \prod_{\nu < \mu} |\lambda_{\nu} - \lambda_{\mu}| \delta \left(\sum_{\nu} \lambda_{\nu} - M\right)
\]
(using $\Tr(S) = M$), which is the fixed trace JPDF for the Wishart/Laguerre ensemble.

\section{Conclusions}\label{conclusions}

In summary, we have shown how, in the limit of large matrix size, discrete random walks through the space of various Bernoulli ensembles lead  to a stochastic motion that converges to the well-known Dyson Brownian motion model for the classical Gaussian ensembles. The key difference in our approach is that we do not assume any Brownian motion of the matrix elements from the outset, instead this motion arises simply from changing the sign of the matrix elements, randomly one by one. By calculating the moments of this evolution we obtain an approximate Fokker-Planck equation (\ref{FP eqn main}), and show the error decreases as the matrix dimension becomes larger. If one were to neglect this error then stationary solution would correspond to JPDF of certain well known fixed trace ensembles with Gaussian elements.

We also believe that the methods could be adapted to treat ensembles in which the matrix elements are chosen from probability distributions other than Bernoulli. If these distributions are sufficiently well-behaved then again, we would expect similar results based upon the ideas of the CLT mentioned in the paper.

During this work we have relied on the results of previous authors regarding the delocalisation of eigenvectors and the rigidity, or gaps, in eigenvalues. These were necessary in order to justify neglecting higher order terms in the perturbation theory, as well as higher moments stemming from the Taylor expansion (\ref{FP expansion new}). Since we would like to focus on the eigenvalues, it would be interesting to see whether such necessities can be circumnavigated. In fact, in a subsequent paper \cite{Joyner-2015} we show how this can be achieved for similar unconstrained Bernoulli ensembles, as appear here. The approach is to develop a Fokker-Planck description for the traces $t_n=\sum_{\nu=1}^N \lambda_{\nu}^n, \ \ \ 1\le n\le N$ and then perform a transformation of variables back to the eigenvalues. The advantage of this method is that, in contrast to the approach in the present paper, it is non-perturbative. It may therefore offer a way to treat constrained random graph ensembles such as the random regular tournaments and random regular graphs outlined in the introduction, for which presently current methods cannot be applied.

The methods may also be applicable to the spectral statistics of other graph operators, such as the graph Laplacian $L=-A+D$ (where $D$ is a diagonal matrix of the vertex degrees $d_i = \sum_j A_{ij}$) or discrete Schr\"odinger operators $H=L+V$ (where $V$ is a random diagonal potential). These offer interesting alternatives to the matrices considered here, since one observes differences in the spectral properties, such as the mean spectral density \cite{Ding-2010} or the localisation of eigenvectors \cite{Metz-2010,Slanina-2012}.

\vspace{10pt}

\noindent \textbf{Acknowledgements:} The authors would like to thank M. Aizenman, E. Paquette and P. Sosoe for many useful and enlightening discussions, M. Nowak for recommending references pertaining to the Wishart ensemble and S. Gnutzmann for offering numerous critical comments and suggestions to the text. Support from the Israel Science Foundation grant ISF - 861/11 (F.I.R.S.T.)  is acknowledged and CJ would also like to thank the Feinberg Graduate School for financial support.

\appendix

\section{Higher moments}\label{Higher moments}

In the following we estimate the leading contributions coming from the third and fourth moments. We do this because these are odd and even moments and so their contributions will be slightly different. All others are of a similar suit and therefore we do not include the calculations.

\subsection{Cubic terms}\label{App: Cubic terms}

The linear and quadratic components provide us with the leading order contributions to the first and second moments, however we must also estimate higher moments. This means calculating the leading contributions to the moments $M_{\nu\mu\kappa}$, $M_{\nu\mu\kappa\gamma}$ etc., which will allow us to provide error estimates in our Fokker-Planck equation. In addition, as we shall discover, the main correction to $M_{\nu}$ will come, not from the errors arising in the calculations of the linear and quadratic terms, but from contributions coming from higher orders in the perturbation expansion (\ref{Perturbation formula}). The third term of which is given by
\begin{equation}\label{Third order perturbation}
\fl
\sum_{\kappa \neq \nu}\sum_{\mu \neq \nu} \frac{\langle \nu | \Delta \bar{B} | \mu \rangle \langle \mu \Delta \bar{B} | \kappa \rangle \langle \kappa | \Delta \bar{B} | \nu \rangle}{(\lambda_{\nu} - \lambda_{\mu})(\lambda_{\nu} - \lambda_{\kappa})} +
\sum_{\mu \neq \nu} \frac{|\langle \nu | \Delta \bar{B} | \mu \rangle|^2\langle \nu | \Delta \bar{B} | \nu \rangle}{(\lambda_{\nu} - \lambda_{\mu})^2}.
\end{equation}
This requires us to estimate quantities of the form
\begin{equation}\label{Cubic expectation}
\E(\langle \nu |\Delta \bar{B}| \mu \rangle \langle \mu |\Delta \bar{B}| \kappa \rangle \langle \kappa |\Delta \bar{B}| \nu \rangle).
\end{equation}
Once again, we must follow the same procedure as before by summing over paths $\omega$, however the split into diagonal and off-diagonal terms is not quite so simple. The basic idea goes back to the fact that we have a sum over (almost independent) random variables, $\Delta \bar{B}^{\omega} = \sum_r \delta \bar{B}^{\omega_r}$ and that terms of the type $\E(|\langle \nu |\delta \bar{B}| \mu \rangle|^2)$ will give a larger contribution than the $|\E(\langle \nu |\delta \bar{B}| \mu \rangle)|^2$. Therefore when $\nu = \mu$ for instance we have that (\ref{Cubic expectation}) is approximately equal to
\[
\Delta t^2 \E(\langle \nu |\delta \bar{B}| \nu \rangle)\E( |\langle \nu |\delta \bar{B}| \kappa \rangle|^2) + \mbox{corrections},
\]
which gives a leading contribution $\Delta \eta \O{c-3}$ (if all three $\nu$, $\mu$ and $\kappa$ were different then it would be 0). However we know from \cite{Tao-2011b} the denominator in (\ref{Third order perturbation}) is of order $\O{-2 -q}$ with a probability that tends to 1 in the limit of large $N$ (see Section \ref{Spectral statistics}. Therefore we get a correction to $M_{\nu}$ of order $\O{c+q-1}$, which is in fact the dominant error to this time-averaged moment. A more detailed calculation of (\ref{Cubic expectation}) begins with summing over all relevant paths and expanding $\Delta \bar{B}^{\omega}$ in terms of components $\omega_r \in \omega$, just as we have done before in Section \ref{Subsec: Linear terms} and Section \ref{Subsec: Quadratic terms}, i.e.
\[
\sum_{X = 0}^{\Delta t}\frac{P_{\Delta t}(X)}{|\Omega_X|}\sum_{\omega \in \Omega_X} \sum_{r,s,u}\langle \nu |\delta \bar{B}^{\omega_r}| \mu \rangle \langle \mu |\delta \bar{B}^{\omega_s}| \kappa \rangle \langle \kappa |\delta \bar{B}^{\omega_s}| \nu \rangle.
\]
This will give us contributions coming from paths $\omega$ that contain either one, two or three elements that are the same. So we will get terms of the type
\[
\fl
\sum_{X = 0}^{\Delta t}\frac{P_{\Delta t}(X)}{|\Omega_X|}\left[ \Phi^{(1)}_X \sum_e f(e,e,e) +  \Phi^{(2)}_X\sum'_{e_1,e_2} f(e_1,e_1,e_2) + \Phi^{(3)}_X \sum'_{e_1,e_2,e_3}f(e_1,e_2,e_3)\right],
\]
where
\[
f(e_1,e_2,e_3) = \langle \nu |\delta \bar{B}^{e_1}| \mu \rangle \langle \mu |\delta \bar{B}^{e_2}| \kappa \rangle \langle \kappa |\delta \bar{B}^{e_3}| \nu \rangle
\]
and as before $e_i \in \{pq\}_{p\leq q}$. We emphasise this is not exact but suffices to illustrate how the main contributions arise. In particular (comparing to equation (\ref{Quadratic expansion})) the primed summands represent summations in which all the variables are not equal to each other. Though removing this restriction will only produce negligible errors. Moreover, we will have additional terms of the form $f(e_1,e_2,e_1)$, $f(e_2,e_1,e_1)$ etc. but these will be of the same order as equivalent terms (i.e. $f(e_1,e_1,e_2)$) and so since these are finite in number will only lead to some constant, independent of $N$, multiplying the contribution from $f(e_1,e_1,e_2)$, which we neglect anyway. The three terms give respective contributions
\begin{equation}\label{Cubic term 1}
\fl
\mathcal{O}(\Delta \eta) \sum_{p,q} \bar{B}^3_{pq} (\nu_p\mu_q + \nu_q\mu_p)(\mu_p\kappa_q + \kappa_q\mu_p)(\kappa_p\nu_q + \kappa_q\nu_p).
\end{equation}
\begin{equation}\label{Cubic term 2}
\fl
\mathcal{O}(\Delta \eta^2) \left(\sum_{p,q} \bar{B}_{pq} (\nu_p\mu_q + \nu_q\mu_p)\right) \left(\sum_{p,q} \bar{B}_{pq}^2(\mu_p\kappa_q + \kappa_q\mu_p)(\kappa_p\nu_q + \kappa_q\nu_p)\right)
\end{equation}
and
\begin{equation}\label{Cubic term 3}
\fl
\mathcal{O}(\Delta \eta^3)\left( \sum_{p,q} \bar{B}_{pq} (\nu_p\mu_q + \nu_q\mu_p)\right) \left(\sum_{p,q} \bar{B}_{pq}(\mu_p\kappa_q + \kappa_q\mu_p) \right) \left( \sum_{p,q}\bar{B}_{pq}(\kappa_p\nu_q + \kappa_q\nu_p)\right).
\end{equation}
where again we have replaced the summation $\sum_{p \leq q}$ with $\sum_{p,q}$ at the expense of a negligible error. Let us now estimate each term. For (\ref{Cubic term 1}), if we have two equal eigenvalues, e.g. $\nu = \mu$, then it becomes
\[
\fl
\Delta \eta \mathcal{O}(N^{-1})\sum_{p,q} 2\bar{B}_{pq}\nu_p\nu_q(\nu_p\kappa_q + \nu_q\kappa_p)^2 \approx \Delta \eta \mathcal{O}(N^{-3})\sum_{p,q} \bar{B}_{pq}[\nu_p\nu_q + \kappa_p\kappa_q],
\]
where we have assumed again that the eigenvectors are delocalised, meaning that we take $\nu_p^2$ and $\kappa_p^2 = \O{-1}$. Since the summation is of order unity, we can estimate (\ref{Cubic term 1}) to be $\O{-3}\Delta \eta$. From the third term in the perturbation expansion (\ref{Third order perturbation}) we see that at most two of $\nu$, $\mu$ and $\kappa$ may be the same. If all three are different then (\ref{Cubic term 2}) is 0 due to the first summation, however if we have $\nu = \mu$ then (\ref{Cubic term 2}) is equal to
\[
\mathcal{O}(\Delta \eta^2) \frac{\lambda_{\nu}}{N} = \O{c - 3}\Delta \eta.
\]
Finally one can see the contribution (\ref{Cubic term 3}) is 0 (up to subleading corrections of course). Thus we see that the dominant correction comes from (\ref{Cubic term 2}), which is what one would expect from the central limit theorem, as it is quadratic in the time difference $\Delta t$.

\vspace{20pt}
We now turn our attention to quantities of the form
\[
\E(\langle \nu |\Delta \bar{B}| \nu \rangle \langle \mu |\Delta \bar{B}| \mu \rangle \langle \kappa |\Delta \bar{B}| \kappa \rangle).
\]
These provide the leading order contributions to the third order moments $M_{\nu\mu\kappa}$ and are of a slightly different to (\ref{Cubic expectation}). Nevertheless they also give the same order contribution in $N$. To see this we proceed in exactly the same as before to obtain contributions of the type
\begin{equation}\label{Cubic term 1b}
\mathcal{O}(\Delta \eta) \sum_{p,q} \bar{B}^3_{pq} \nu_p\nu_q \mu_p\mu_q \kappa_p\kappa_q
\end{equation}
\begin{equation}\label{Cubic term 2b}
\mathcal{O}(\Delta \eta^2) \left(\sum_{p,q} \bar{B}_{pq}\nu_p\nu_q \right) \left(\sum_{p,q} \bar{B}_{pq}^2 \mu_p\mu_q \kappa_p\kappa_q\right)
\end{equation}
and
\begin{equation}\label{Cubic term 3b}
\mathcal{O}(\Delta \eta^3) \left(\sum_{p,q} \bar{B}_{pq}\nu_p\nu_q \right) \left(\sum_{p,q} \bar{B}_{pq} \mu_p\mu_q\right)\left(\sum_{p,q} \bar{B}_{pq} \kappa_p\kappa_q\right).
\end{equation}
Once again, the linear time contribution (\ref{Cubic term 1b}) is a fluctuating quantity that we estimate to be very small. The cubic contribution is given by $\Delta \eta \O{2c-4}\lambda_\nu\lambda_{\mu}\lambda_{\kappa}$ and is subdominant. This leaves (\ref{Cubic term 2b}). However, as we have already seen in Section \ref{Subsec: Quadratic terms}, this involves $\sum_{p,q} \bar{B}_{pq}^2 \mu_p\mu_q \kappa_p\kappa_q = \delta_{\mu\kappa}/N$. Therefore if $\mu=\kappa$ this gives a contribution $\O{c - 3}\Delta \eta$ and so applies for all terms of the type $M_{\nu\mu\mu}$, in which two or more of the eigenvalues are the same.

\subsection{Quartic terms}\label{App: Quartic terms}

The evaluation of quartic terms follows exactly the same line of reasoning. They appear in two forms, either
\begin{equation}\label{Quartic first type}
\E(\langle \nu |\Delta \bar{B}| \mu \rangle \langle \mu |\Delta \bar{B}| \kappa \rangle \langle \kappa |\Delta \bar{B}| \gamma \rangle  \langle \gamma |\Delta \bar{B}| \nu \rangle),
\end{equation}
which appear in the higher order terms of the perturbation expansion (\ref{Perturbation formula}), or as the leading contributions to the moments $M_{\nu\mu\kappa\gamma}$, for which they take the form
\[
\E(\langle \nu |\Delta \bar{B}| \nu \rangle \langle \mu |\Delta \bar{B}| \mu \rangle \langle \kappa |\Delta \bar{B}| \kappa \rangle  \langle \gamma |\Delta \bar{B}| \gamma \rangle).
\]
Let us briefly outline the latter case. Suppose that all parameters are equal, as is the case for the moments $M_{\nu\nu\nu\nu}$. If we compare with the expressions (\ref{Cubic term 1b}), (\ref{Cubic term 2b}) and (\ref{Cubic term 3b}) then we see that any contributions with a factor $(\frac{1}{d_N}\sum_{p,q}\bar{B}_{pq}\nu_p\nu_q) = \lambda_{\nu}/d_N$ will give a subleading contribution. In other words we have that $\E(\langle \nu | \delta \bar{B} | \nu \rangle^2)$  is of a higher order in $N$ than $\E(\langle \nu | \delta \bar{B} | \nu \rangle)^2$. This leaves us with two potential contributions. The first comes from the diagonal term, which is
\[
\mathcal{O}(\Delta \eta) \sum_{p,q} \bar{B}_{pq}^4 \nu_p^4\nu_q^4 = \Delta \eta\O{-4}.
\]
The second comes from
\[
\mathcal{O}(\Delta \eta^2) \left(\sum_{p,q} \bar{B}_{pq}^2 \nu_p^2\nu_q^2\right)^2 = \Delta \eta\O{c-4}.
\]
Thus we see the latter is of a larger order in $N$ since $c>0$. This will also be the case when we have two pairs of equal eigenvalues, i.e. moments of the type $M_{\nu\nu\mu\mu}$ are also of the same order, however this is not the case when at least 3 eigenvalues are different and so moments of the type $M_{\nu\mu\kappa\kappa}$ for instance are sub-leading.

In addition, we may also find terms of the first type (\ref{Quartic first type}) which exhibit also contribution of order $\Delta \eta\O{c-4}$. This means they provide an error to $M_{\nu}$ approximately $\O{c +3s -1}$, since terms in the denominator of the fourth term in the perturbation expansion are of the type $(\lambda_{\nu} - \lambda_{\mu})^3 = \O{-3-3q}$. Therefore we see that fourth order terms give a marginally larger contribution to $M_{\nu}$ than third order terms, which contribute $\O{c + 2q - 1}$. However, beyond this we find $\O{2c + 4q - 2}$ and $\O{2c + 5q - 2}$ for the fifth and sixth orders, which are subleading in comparison to the fourth.

\end{document}